\documentclass[12pt]{article}
\usepackage{graphicx}
\usepackage[font=footnotesize]{caption}
\usepackage{natbib}
\usepackage{fancyhdr,amsmath,amssymb, amsthm,mathtools}
\usepackage{etoolbox}

\usepackage[colorlinks,citecolor=blue,urlcolor=blue]{hyperref}
\usepackage{booktabs}
\usepackage{tikz}
\usetikzlibrary{fit,positioning,calc}
\usepackage{scrextend}
\usepackage{hyperref}
\usepackage{chngcntr}
\usepackage{apptools}
\usepackage{bbm}
\usepackage{multirow}
\usepackage[T1]{fontenc}
\usepackage[export]{adjustbox}
\usepackage{mathrsfs}
\usepackage{mathrsfs,dsfont}
\usepackage{amsfonts}
\usepackage{stmaryrd}
\usepackage{mathrsfs,dsfont}
\usepackage{bbm,bm}
\usepackage{algorithm}
\usepackage{algpseudocode}
\usepackage{setspace}
\usepackage{subfigure}
\usepackage{upgreek}

\RequirePackage{xr-hyper}

\allowdisplaybreaks

\AtAppendix{\counterwithin{Lemma}{section}}

\def\b0{\mathbf{0}}

\def\ba{\mathbf{a}}

\def\bw{\mathbf{w}}

\def\bz{\mathbf{z}}

\newcommand{\dBeta}{\mathrm{Beta}}

\newcommand{\dCat}{\mathrm{Cat}}

\def\bP{\mathbf{P}}

\def\bY{\mathbf{Y}}

\def\balpha{\boldsymbol{\alpha}}

\def\bgamma{\boldsymbol{\gamma}}

\def\btheta{\boldsymbol{\theta}}

\newcommand{\id}{\mathds{1}}

\def\pr{\textrm{\textup{pr}}}

\newcommand{\appropto}{\mathrel{\vcenter{
  \offinterlineskip\halign{\hfil$##$\cr
    \propto\cr\noalign{\kern2pt}\sim\cr\noalign{\kern-2pt}}}}}

\definecolor{green}{rgb}{0.3, 0.73, 0.09}

\newcommand{\magenta}[1]{\!}

\makeatletter
\providecommand{\leftsquigarrow}{%
  \mathrel{\mathpalette\reflect@squig\relax}%
}
\newcommand{\reflect@squig}[2]{%
  \reflectbox{$\m@th#1\rightsquigarrow$}%
}
\makeatother

\makeatletter
\newcommand\subsubsubsection{\@startsection{paragraph}{4}{\z@}{-2.5ex\@plus -1ex \@minus -.25ex}{1.25ex \@plus .25ex}{\normalfont\normalsize\bfseries}}
\makeatother

\newcommand{\blind}{1}

\expandafter\def\expandafter\normalsize\expandafter{%
    \normalsize%
    \setlength\abovedisplayskip{11pt}%
    \setlength\belowdisplayskip{11pt}%
    \setlength\abovedisplayshortskip{11pt}%
    \setlength\belowdisplayshortskip{11pt}%
}

\begin{document}

\def\spacingset#1{\renewcommand{\baselinestretch}%
{#1}\small\normalsize} \spacingset{1}

\if1\blind
{
	\title{\vspace{-28pt}
	\Large{Dependent Stochastic Block Models for \\ Age--Indexed Sequences of Directed Causes--of--Death Networks}}
\author{Giovanni Roman\`o$^{\dagger}$, Cristian Castiglione$^{\star}$ and Daniele Durante$^{\star, \dagger}$\\
$^{\star}$Bocconi Institute for Data Science and Analytics, Bocconi University\hspace{.2cm}\\
	 $^{\dagger}$Department of Decision Sciences, Bocconi University}
	\date{}
	\maketitle
} \fi

\if0\blind
{
	\bigskip
	\bigskip
	\bigskip
	\begin{center}
		{\LARGE\bf Dependent Stochastic Block Models for Age--Indexed Sequences of Directed Causes--of--Death Networks}
	\end{center}
	\medskip
} \fi

\vspace{-23pt}

\begin{abstract}

Death events commonly arise from complex interactions among interrelated causes, formally classified in reporting practices as {\em underlying} and {\em  contributing}. Leveraging information from death certificates, these interactions can be naturally represented through a sequence of directed networks encoding co--occurrence strengths between pairs of underlying and contributing causes across ages. Although this perspective opens the avenues to learn 
informative age--specific block interactions among endogenous groups of underlying and contributing causes displaying similar co--occurrence patterns, there has been limited research along this direction in mortality modeling. This is mainly due to the lack of suitable stochastic block models for sequences of directed networks indexed by a predictor. We cover this gap through a novel Bayesian formulation which crucially learns two separate group structures for underlying and contributing causes, while allowing both structures to change smoothly across ages via dependent random partition priors. As illustrated in simulation studies, the proposed formulation outperforms state--of--the--art solutions that could be adapted to our motivating application. Moreover, when applied to \textsc{usa} mortality data, it unveils structures in the composition, evolution, and modular interactions among causes--of--death groups that were hidden to classical demographic studies. Such findings could have relevant policy implications and contribute to an improved understanding of the recent ``death of despair'' phenomena in the \textsc{usa}.

\end{abstract}
\vspace{-2pt}
\noindent%
{\it Keywords:}  Causes of Death; Dependent Random Partition Prior;  Directed Network; Mortality Modeling; Stochastic Block Model 

\spacingset{1.78} 
\vspace{-15pt}

\section{Introduction}\label{sec_1}
\vspace{-10pt}
Although demographic research has traditionally focused on investigating mortality phenomena via overall longevity indicators 
\citep[e.g.,][]{canudas2010three,van2021have}, recent state--of--the--art~studies have suggested that a comprehensive understanding of the core determinants behind modern mortality trends necessarily requires a finer--scale analysis disaggregating such trends across causes of death  \citep[e.g.,][]{egidi2018network,woolf2019life,canudas2020reflection,bergeron2020diversification,mehta2020us,grippo2020multi,stefanucci2022analysing,trias2023cause,calazans2023levels}. Besides yielding a deeper understanding of mortality patterns, these analyses are also of paramount importance to devise innovative policies in public health, and evaluate the corresponding effects across multiple, often interrelated, causes of death \citep[e.g.,][]{aburto2018potential,bergeron2020diversification}.

\vspace{-1pt}

The relevance of the above endeavor combined with the availability of increasingly--refined data~resources (see, e.g., the “\textsc{who} Mortality Database”, the “Human Causes of Death Data Series“,~the~“Global Burden of Disease“, and the “\textsc{us} National Center for Health Statistics“), have produced an unprecedented understanding of modern mortality trends that lacked a clear explanation under classical studies focused on aggregated longevity indicators. These advancements have been achieved through the design of inference and predictive methods for the compositions, determinants, diversification and trends of either  the {\em underlying} cause of death (i.e., “the disease or injury which initiated the chain of  events leading directly to death“ \citep{world2004international}) \citep[see, e.g.,][]{foreman2018forecasting,bergeron2020diversification,mehta2020us,stefanucci2022analysing,depaoli2024functional,ahmad2024leading,huynh2024joint} or multiple causes of death~(\textsc{mcod}),~comprising~both the {\em underlying} one and its {\em contributing} causes (i.e.,  “all other significant conditions contributing to death but not resulting in the underlying cause“  \citep{world2004international}) \citep[see, e.g.,][]{desesquelles2010revisiting,desesquelles2012analysing,desesquelles2014cause,moreno2017survival,egidi2018network,grippo2020multi,trias2023cause,bishop2023analysis,grippo2024beyond}. Recalling the comprehensive review by \citet{bishop2023analysis}, among these two perspectives, the second has been object of increasing interest in the recent years since it aligns more closely with the fact that death events are commonly associated with complex systems of multiple interrelated causes, rather than a single one in isolation \citep[e.g.,][]{israel1986analytical,redelings2006comparison,desesquelles2014cause,trias2023cause}. As a consequence, \textsc{mcod} analyses have potential to provide more refined and realistic understanding of modern mortality patterns, while opening the avenues to study  the higher--level system of  relational structures among underlying and contributing causes, along with its changes across~age~classes. 

\vspace{-2pt}

Leveraging the available data from death certificates, the aforementioned relational system can be naturally represented via a sequence of  networks whose directed edges measure, for each~age~class,~the strength of the co--occurrence relation ``cause $i$ appears as the underlying of the contributing cause~$j$''. However, despite its potential in unveiling yet--unexplored relational structures among causes of death with promising policy impact \citep[e.g.,][]{desesquelles2010revisiting,desesquelles2012analysing,desesquelles2014cause}, this network perspective~has been mostly overlooked in  \textsc{mcod} studies. As discussed in the following, a key barrier toward advancing along this direction can~be~found in the lack of suitable statistical models capable of uncovering relevant and interpretable structures~that drive the formation and evolution of the complex co--occurrence~patterns among underlying and contributing causes of death, across~ages. In fact, similarly to epidemiology studies of co--morbidity networks \citep[e.g.,][]{jeong2017network,fotouhi2018statistical,jones2023methods}, current network--based~\textsc{mcod} analyses \citep[e.g.,][]{egidi2018network,ukolova2023racial} rely on traditional summary measures~applied to simplified versions of the original data, which specialize the analysis to a single age~class~and~do not consider the distinction between underlying and contributing causes. Therefore, although these contributions have~the~merit of showcasing the potential~of~the~network~perspective within \textsc{mcod} studies, the resulting findings are descriptive in nature~and~do~not~inform on how  relational structures among underlying and contributing~causes~vary~across~ages.

\vspace{-2pt}

As showcased in our application to \textsc{usa} mortality data in 2019 (see Sections~\ref{sec_1.D} and \ref{sec_5}), overcoming the above limits necessarily requires a model--based perspective capable of accounting for the full complexity~of causes--of--death co--occurrence patterns across several dimensions, while quantifying uncertainty in~the inferred structures behind these observed patterns. Advancements along these lines are not only essential to possibly refine current health care investment policies in a phase characterized by a higher diversification, increasing complexity and lower predictability of multiple causes--of--death landscapes \citep[see, e.g.,][]{bergeron2020diversification,trias2023cause}, but could also help in resolving recent debates on the determinants behind modern mortality trends. A relevant one, which motivates our focus on \textsc{usa} causes--of--death data, revolves around the recent stagnation in the \textsc{usa} life expectancy \citep[e.g.,][]{woolf2019life,mehta2020us,case2021deaths}. While such a stagnation has been attributed to a growing mid--life mortality, the determinants underlying this mortality increment have generated a debate around different views that either support causes such as drug overdoses, alcohol abuses and suicides (``death of despair'') \citep[e.g.,][]{woolf2019life,case2021deaths} or identify cardiovascular diseases as main drivers \citep[e.g.,][]{mehta2020us}. Albeit different, both views arise from the study of underlying causes. Hence,~as illustrated in Section~\ref{sec_5}, the current debate can find a consensus under a refined model--based analysis of causes--of--death networks that is capable of inferring age--specific  block interactions~among~endogenous groups of underlying and contributing causes displaying similar co--occurrence patterns. These~inferred clustering structures and the associated block interactions might, in fact, unveil unexplored modules  among seemingly unrelated causes that were object of past debates. In addition, as illustrated within Figure~\ref{fig:intro:data_plot}, such a perspective~yields an interpretable reconstruction of complex modules in causes--of--death networks that unveil within-- and across--group diversification structures in both underlying and contributing causes, along with the associated changes at different age classes. This is a key to shift~the focus of current investment policies away from targeting a single underlying cause~with~a~high prevalence in the population and toward jointly prioritizing internally--homogenous~groups~of~underlying~and contributing causes characterized by remarkable modular co--occurrences at given age classes.

\vspace{-1pt}

Motivated by the above endeavor, we generalize stochastic block model (\textsc{sbm}) representations~\citep[e.g.,][]{holland1983stochastic,nowicki2001estimation}  to learn informative structures in age--indexed sequences of categorically--weighted directed networks among underlying and contributing causes~of~death. As clarified in Sections~\ref{sec_2}--\ref{sec_3}, the proposed formulation (i) learns two separate group structures for underlying and contributing causes, respectively, (ii) allows these structures to change smoothly across age classes via dependent random partition priors \citep{page2022dependent} further informed by external macro–classifications of death causes through product partition models \citep{muller2011product}, (iii) automatically estimates the number of groups for each network in the sequence, (iv) accounts for flexible block interactions among these inferred groups and (v) facilitates principled uncertainty quantification and inclusion of expert knowledge under a Bayesian approach to inference. 

\vspace{-1pt}

Although classical \textsc{sbm}s have witnessed effective extensions in several directions over the recent years, a flexible formulation addressing (i)--(v) within a single construction is lacking in the literature. In fact, while  state--of--the--art dynamic stochastic block models could be possibly adapted~to~our~motivating application by replacing time with age, available formulations \citep[e.g.,][]{ishiguro2010dynamic,yang2011detecting,xu2014dynamic,xu2015stochastic,matias2017statistical,pensky2019spectral,goto2021clustering} are not designed to infer two separate partitions for the rows and columns of the adjacency matrices characterizing the observed directed networks. Furthermore, these models lack strategies to inform such grouping structures by external node attributes, often focus on binary edges, and generally cannot learn automatically the number of groups for each network in the sequence.~While~these~issues~have~been addressed separately in the literature \citep[e.g.,][]{tallberg2004bayesian,kemp_2006,mariadassou2010uncovering,rohe2016co,zhang2016community,zhang2022directed,geng_2019,legramanti2022extended,gaffi2025partially},~the overarching focus of available contributions has been on static, single--network, settings, rather than~on sequences of networks indexed~by~an~ordered~covariate~(e.g.,~time~or~age).

The  importance of covering the above methodological gap is illustrated through realistic simulation studies in Section~\ref{sec_4}, where the model we propose in Sections~\ref{sec_2}--\ref{sec_3}  is shown to outperform~and~extend the inference potential of state--of--the--art dynamic \textsc{sbm}s~that could be possibly employed in the applied settings motivating our contribution. These advantages are strengthened in the application to \textsc{usa} mortality data within Section~\ref{sec_5}. In this case, our model unveils yet--unexplored~structures in the composition, evolution, diversification and modular interactions among underlying and contributing causes--of--death groups that were hidden to previous demographic studies. Such findings may have important policy implications and~open~the~avenues~to~achieve a more comprehensive understanding of the determinants underlying the recent stagnation in the \textsc{usa} life expectancy \citep[e.g.,][]{woolf2019life,mehta2020us,case2021deaths}. Concluding remarks~can~be~found~in~Section~\ref{sec_6}.

\vspace{-5pt}

\subsection{USA Causes--of--Death Data}\label{sec_1.D}
\vspace{-4pt}
As anticipated in Section~\ref{sec_1}, our motivating application arises from the attempt to unveil higher--level and more nuanced determinants behind the alarming stagnation in the \textsc{usa} life expectancy \citep[e.g.,][]{woolf2019life,mehta2020us,case2021deaths} through a novel perspective that avoids overly--simplified analyses of the underlying cause of death, but rather~studies~the~complex~interaction systems among multiple, interrelated, causes.

To this end, we focus on age--indexed sequences of co--occurrence networks among underlying and contributing causes extracted from the $\approx 2{,}860{,}000$ death certificates recorded in the  \textsc{usa} for 2019 (see \url{https://wonder.cdc.gov/mcd-icd10.html}). Such certificates are issued as part of the National~Vital Statistics System maintained by the National Center for Health Statistics (\textsc{nchs}), and document, for each death event, key demographic information, such as gender and age, alongside a single underlying cause and its contributing ones. All these causes are identified according to the International~Classification of Diseases (\textsc{icd}) system (now in its 11th revision), which employs different levels of granularity, ranging from a single alphabetic character associated with broad macro--categories, to the finest--scale seven--character code that identifies highly--detailed sub--categorizations for each cause. Such a latter classification yields thousands of mortality factors, with a large portion comprising  extremely--specific and highly--rare causes. For this reason, and consistent with our overarching focus, we consider the classification based on the first two characters of the \textsc{icd} codes, upon discarding never--observed causes and those not related to diseases and pre--existing medical conditions (i.e., injuries and external). This choice yields a total of $n=139$ causes under analysis,~thereby~achieving~an~effective~balance between overly--aggregated studies that fail to unveil nuanced patterns, and excessively--fine~classifications which do not facilitate interpretable analyses and may be more prone to reporting issues due to increased challenges in disentangling highly--similar causes \citep[e.g.,][]{redelings2007using,desesquelles2010revisiting,desesquelles2012analysing,desesquelles2014cause,bishop2023analysis}. Leveraging routine practice in demographic studies \citep[see, e.g.,][]{desesquelles2010revisiting,lozano2012global,trias2023cause,depaoli2024functional}, we further stratify the data by age classes defined as $[0,1]$, $(1,10]$, $(10,20]$, $\dots$, $(90,100]$, where the first group accounts for the peculiar patterns associated with infant death events.

\vspace{-2pt}

Leveraging the above stratifications of the death certificates together~with~the~causes--of--death~classification considered, it is thus possible to record, for each age class $x$, the total number of~certificates that have cause $i$ as the underlying of the contributing $j$,~for~every  $i=1, \ldots, n$~and $j=1, \ldots, n$. While these pairwise counts already yield a sequence of causes--of--death co--occurrence~networks across ages, we opt for analyzing a discretized version of such directed networks which classifies~the~original counts into four interpretable levels, defined as  \texttt{absent} (0), \texttt{rare} (1--10), \texttt{present} (11--100), and \texttt{frequent} ($>$100). This choice is motivated by two main reasons.~First,~it~facilitates~interpretation by providing policy experts with findings based on a simple~and~intuitive~four--level measure of co--occurrence strengths among causes of death. Second, it is beneficial in reducing~the~noise and partial distortions that may arise from the aggregation of hundreds of thousands of death certificates compiled by different specialists across \textsc{usa} in a year.  Note that, although~different thresholds may be considered, those we identify are based on the analysis of the empirical distribution of the co--occurrence~counts, whose median across ages and causes--of--death pairs is $\approx 15$.~This~motivates our focus on multiples of 10, which proved also robust when considering other connectivity measures among causes of death that normalize the pairwise counts with respect~to the degree of appearance of the associated causes within the certificates \citep[e.g.,][]{hidalgo2009dynamic,chmiel2014spreading,fotouhi2018statistical};~see~Section~\ref{sec_6}.

\begin{figure}[t!]
\centering
    \includegraphics[trim=0cm 0cm 0cm 0cm,clip,width=0.95\textwidth]{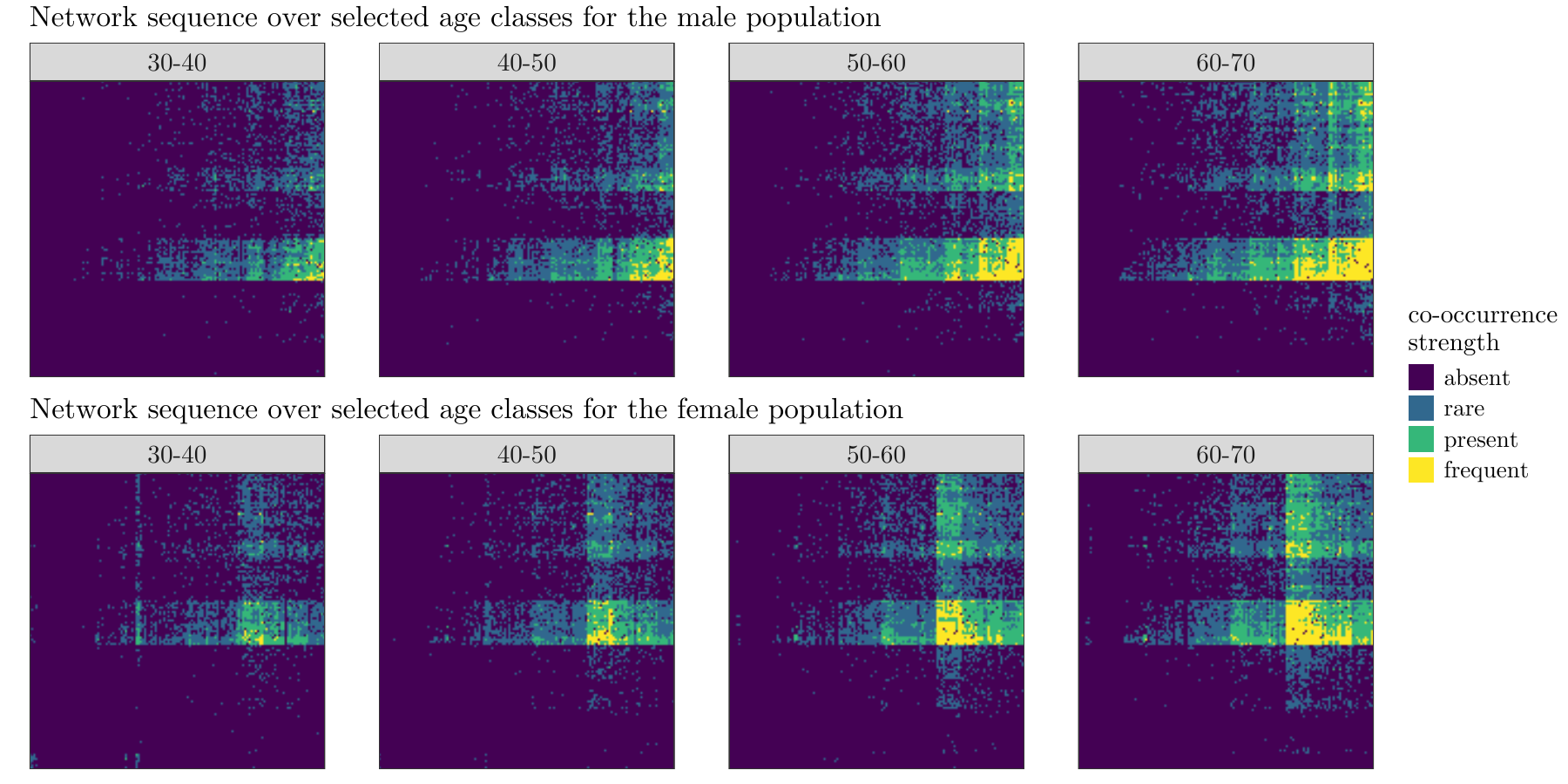}
      \vspace{-3pt}
    \caption{\footnotesize{For the male and female populations, graphical representations of the adjacency matrices associated~with the cause--of--death networks studied, at four selected consecutive age classes. In each matrix, rows (underlying~causes) and columns (contributing causes) are re--ordered according to the two group structures identified by a basic hierarchical clustering algorithm. Such an algorithm is applied separately to the two dimensions of each adjacency matrix leveraging the Hamming distance~\citep{hamming1950error} as a measure of dissimilarity among rows and between columns, respectively.
  \vspace{0pt}}}
    \label{figure:1}
\end{figure}

Figure~\ref{figure:1} illustrates the adjacency matrices associated with the resulting networks under analysis,~at selected age classes, for both the male and female populations. To provide preliminary quantitative~evidence that supports the model developed in Sections~\ref{sec_2}--\ref{sec_3}, the entries of each matrix are re--ordered  to highlight the row and column groups learned by a basic hierarchical clustering algorithm (see the \texttt{R} library \texttt{pheatmap)}.  This algorithm is applied separately to the two dimensions~of~each adjacency matrix, corresponding, respectively, to  underlying (rows) and contributing (columns) causes, and employs the Hamming distance~\citep{hamming1950error} as a measure of dissimilarity among the rows~and between the columns, respectively. As shown in Figure~\ref{figure:1}, there is a clear evidence of block co--occurrence structures in the directed causes--of--death networks under analysis.  These blocks are induced by group patterns among rows and between columns that cannot be assumed equal, but rather should be studied as two separate partitions. Furthermore, albeit estimated separately for each age class, such partitions seem to vary smoothly across consecutive ages in the observed networks. 
\vspace{-2pt}

The above preliminary quantitative findings motivate the dependent \textsc{sbm} for age--indexed sequences of directed causes--of--death networks developed in Sections~\ref{sec_2}--\ref{sec_3}. In fact, while the results in Figure~\ref{figure:1} already provide relevant insights, basic hierarchical clustering is not designed to borrow information between row and column partitions across ages, and does not provide a principled model--based representation of the complex joint system of interactions among underlying and contributing causes across age classes. These issues do not facilitate rigorous inference, uncertainty quantification, supervision by meaningful external information and, possibly, assessment~of~the effects of different policy scenarios via simulations from the generative model. Moreover, the results in Figure~\ref{figure:1} are descriptive in nature, and hence, are also more prone to suffer from the possible distortions that arise from reporting practices~of death certificates, an issue which is not specific to our analysis,~but rather common to all causes--of--death studies \citep[see, e.g.,][]{redelings2007using,desesquelles2010revisiting,desesquelles2012analysing,desesquelles2014cause,bishop2023analysis}. Although recent advancements and automated reporting procedures have substantially improved the quality of death certificates, mitigating possible biases through careful data--preprocessing and principled statistical models that account for uncertainty is still important.~Our contribution moves along these lines, while further addressing potential reporting issues through the supervision by biologically--meaningful external classifications~of~the~causes~of~death.~In~this~respect,~it~shall be  emphasized that certain groups inferred by the novel \textsc{sbm} we propose in Sections~\ref{sec_2} and \ref{sec_3} may be also useful in unveiling systematic reporting practices associated with specific underlying and contributing causes at given age classes. As a result, our contribution can also help in improving the understanding of such practices, thereby motivating the design of additional~guidelines in causes--of--death reporting.

\vspace{-10pt}

\section{Dependent Stochastic Block Models for Directed Networks}\label{sec_2}
\vspace{-8pt}
Define the four levels \texttt{absent}, \texttt{rare}, \texttt{present} and \texttt{frequent} introduced in Section~\ref{sec_1.D} through~the~numerical label $c=1, \ldots, 4$. Moreover, recode the age classes $[0,1]$, $(1,10]$, $(10,20]$, $\dots$, $(90,100]$ into the ordered indexes $x=1, 2, \ldots, m$ (with $m=11$ in our application). Then, the networks analyzed are available in the form of a sequence $\bY_1,\ldots, \bY_{m}$ of $n \times n$ categorically--weighted asymmetric adjacency matrices, where the generic $\bY_{x}$ has entries $y_{ijx} = c$ if, at age class $x$, the strength~of~the~directed~co--occurrence relation ``cause $i$ appears as the underlying of the contributing cause~$j$'' is equal to $ c$.

In Section \ref{sec_31}, we generalize state--of--the--art \textsc{sbm}s to define a probabilistic generative mechanism for $\bY_1,\ldots, \bY_{m}$, which addresses points (i)--(v) discussed in Section~\ref{sec_1} via a single formulation. Such a joint model is designed to infer age--specific block interactions~among endogenous groups of underlying and contributing causes displaying similar co--occurrence patterns. To this end, we employ,~and~learn, two separate grouping structures~that~are~allowed to change smoothly across age classes
via dependent random partition priors further informed by macro--classifications~of~death~causes.~Such~priors~are~inspired by contributions of  \citet{page2022dependent} and  \citet{muller2011product}, and are presented~in~Section~\ref{sec_32}.

\vspace{-6pt}
\subsection{Model Formulation}\label{sec_31}
\vspace{-5pt}
Guided by the empirical evidence~in Figure~\ref{figure:1}, we extend available \textsc{sbm}s to infer separate group~structures for the underlying and contributing causes, respectively, from the block patterns displayed~by~the observed~adjacency matrices $\bY_1,\ldots, \bY_{m}$. Recalling Section~\ref{sec_1}, current  \textsc{sbm}s \citep[e.g.,][]{holland1983stochastic,nowicki2001estimation,schmidt_2013,lee2019review,geng_2019,legramanti2022extended} mostly focus on binary undirected networks, thus requiring only~a~single~partition to flexibly characterize the modular architectures in the observed network. Such a single partition is often employed also~in directed settings under the assumption that the rows and columns~of~the~asymmetric adjacency matrix display a shared group structure \citep[see, e.g.,][]{wang1987stochastic,newman2007mixture,mcdaid2013improved,peixoto2022ordered}. Although this perspective simplifies the model, in practice, it may provide an unrealistic characterization of the data generative mechanisms in directed settings, thus raising misspecification issues that yield to biased inferences. In fact, as illustrated in Figure~\ref{figure:1}, it is reasonable to expect that two generic causes of death having similar co--occurrence patterns when treated as underlying, may not display the~same similarities when considered as contributing.~In addition, these co--clustering relations might also vary across different~age~classes.

Consistent with the above discussion we employ, for each age--class index $x=1, \ldots, m$,~two~separate partitions (one for the underlying causes and the other for the contributing ones), whose Cartesian product  yields a block structure on the adjacency matrix $\bY_{x}$ that clusters together, within~each block, pairs of underlying and contributing causes with similar co--occurrence strengths. More specifically, let $\bz_{(1)x} = (z_{(1)1x}, \dots, z_{(1)nx})$ and $\bz_{(2)x} = (z_{(2)1x}, \dots, z_{(2)nx})$ be the group membership vectors associated to the generic row and column partitions $\mathbb{Z}_{(1)x} = \{ \mathbb{Z}_{(1)1x}, \dots, \mathbb{Z}_{(1)H_xx} \}$ and $\mathbb{Z}_{(2)x} = \{ \mathbb{Z}_{(2)1x}, \dots, \mathbb{Z}_{(2)K_xx} \}$, such that $z_{(1)ix} = h$, for $h=1, \ldots, H_x$, if and only if $i \in \mathbb{Z}_{(1)hx}$ and, analogously, $z_{(2)jx} = k$,~for~$k=1, \ldots, K_x$, if and only if $j \in \mathbb{Z}_{(2)kx}$. Then, extending the classical Bernoulli likelihoods for binary edges to the categorically--weighted interactions characterizing the networks under analysis, we assume that $  (y_{ijx} \mid z_{(1)ix} = h, z_{(2)jx} = k, \btheta_{hkx}) \sim \dCat_{1:4}(\btheta_{hkx})$, independently for every $i=1, \ldots, n$, $j=1, \ldots, n$ and $x=1, \ldots, m$, where $\dCat_{1:4}(\btheta_{hkx})$ is a categorical variable indexed by the probability~vector~$\btheta_{hkx}$, with entries $\theta_{hkxc}=\mbox{pr}(y_{ijx}=c \mid z_{(1)ix} = h, z_{(2)jx} = k) \in (0,1)$, for $c=1, \ldots, 4$. Hence,~consistent~with general \textsc{sbm}s, within--block homogeneity is expressed by assuming~that~the~distribution for each entry $y_{ijx}$ of $\bY_{x}$ only depends on the corresponding row and column groups along with the probability~vector characterizing the block associated to such a pair of groups. This means~that the co--occurrence~strengths among all underlying and contributing causes in groups $h$ and $k$, respectively, are  generated from the same categorical distribution, whose parameters~can~change~across~blocks~(i.e.,~pairs~of~groups),~but~not within block, thereby accounting~for~flexible~modular~structures in the observed networks. 

\setlength\abovedisplayskip{7pt}%
    \setlength\belowdisplayskip{8pt}%
    
    \vspace{-3pt}
To complete our Bayesian formulation, we  require priors on the group membership vectors $\bz_{(1)x} $~and $\bz_{(2)x}$, $x=1, \ldots, m$, together with the block--specific parameters $\btheta_{hkx}$,~for $h=1, \ldots, H_x$, $k=1, \ldots, K_x$ and $x=1, \ldots, m$. Extending routine Bayesian \textsc{sbm}s~\citep[e.g.,][]{nowicki2001estimation,mariadassou2010uncovering,legramanti2022extended} from binary to categorical relationships, it is natural to consider  independent $\mbox{Dirichlet}(\ba^\theta = (a_1^\theta, \dots, a_4^\theta))$ priors for the block--specific parameters $\btheta_{hkx}$, $h=1, \ldots, H_x$, $k=1, \ldots, K_x$ and $x=1, \ldots, m$. Besides corresponding to the multivariate generalization of the~Beta priors employed for the block probabilities in binary networks, such a choice achieves conjugacy with the $\dCat_{1:4}(\btheta_{hkx})$ distribution assumed for the entries of $\bY_x$. As such, the block--specific parameters~can be integrated out analytically to obtain a Dirichlet--categorical joint likelihood for $\bY_1,\ldots, \bY_{m}$ conditioned on the group membership vectors $\bz_{(1)x} $ and $\bz_{(2)x}$, $x=1, \ldots, m$. In particular
\begin{eqnarray}\label{eq:model:categ_marg_lik}
p(\bY_1, \ldots, \bY_{m} \mid  (\bz_{(1)x}, \bz_{(2)x}), x=1, \ldots, m)   
    = \prod_{x = 1}^{m}\Bigg(\prod_{h = 1}^{H_x} \prod_{k = 1}^{K_x} \left[ \frac{\Gamma(a_{\bullet}^\theta)}{\Gamma(a_{\bullet}^\theta + n_{hkx\bullet})} \prod_{c = 1}^{4} \frac{\Gamma(a_c^\theta + n_{hkxc})}{\Gamma(a_c^\theta)} \right]\Bigg),
\end{eqnarray}
where $n_{hkxc}$ represents the number of pairs $(i, j)$ such that $z_{(1)ix} = h$, $z_{(2)jx} = k$ and $y_{ijx} = c$,~while $n_{hkx\bullet} = \sum_{c = 1}^{4} n_{hkxc}$ and $a_\bullet^\theta = \sum_{c = 1}^{4} a_c^\theta$. The likelihood in \eqref{eq:model:categ_marg_lik} formally treats the quantities $\btheta_{hkx}$,~$h=1, \ldots, H_x$, $k=1, \ldots, K_x$, $x=1, \ldots, m$ as nuisance parameters and focuses inference on the group structures~encoded~in  $\bz_{(1)x} $ and $\bz_{(2)x}$, for $x=1, \ldots, m$. While learning the block--specific probability~vectors $\btheta_{hkx}$ is also of interest, this perspective is common in \textsc{sbm} formulations \citep[e.g.,][]{wyse2012block,mcdaid2013improved,schmidt_2013,legramanti2022extended,gaffi2025partially}, whose primary interest lies in uncovering groups of nodes that display similar connectivity~patterns within the observed network. In our context, such an information is encoded in  $\bz_{(1)x} $ and $\bz_{(2)x}$, for $x=1, \ldots, m$, thereby motivating our main focus on these two vectors under the Dirichlet--categorical joint likelihood in  \eqref{eq:model:categ_marg_lik}. As discussed in Section~\ref{sec_3}, this choice further facilitates posterior computation, and does not prevent from obtaining ex--post sensible estimates of $\btheta_{hkx}$,~$h=1, \ldots, H_x$, $k=1, \ldots, K_x$,~$x=1, \ldots, m$.

    \vspace{-1pt}

Notice that the likelihood in  \eqref{eq:model:categ_marg_lik} factorizes across the age--class indexes $x=1, \ldots, m$. Although this formulation is useful to avoid overly--sophisticated and intractable representations,~as~shown~within~Figure~\ref{figure:1}, causes--of--death networks exhibit a form of dependence across ages, which is visible in terms of smooth transitions for the corresponding group structures over contiguous age classes. In Section~\ref{sec_32}, we incorporate this dependence through carefully designed priors for the two sequences  $\bz_{(1)1}, \ldots,  \bz_{(1)m}$ and $\bz_{(2)1}, \ldots,  \bz_{(2)m}$ that further include external information on cause--of--death macro--classifications.

\vspace{-9pt}
\subsection{Prior for Dependent Sequences of Random Partitions}\label{sec_32}
\vspace{-6pt}
As anticipated in Section~\ref{sec_31}, we elicit priors for the two sequences $\bz_{(1)1}, \ldots,  \bz_{(1)m}$ and $\bz_{(2)1}, \ldots,  \bz_{(2)m}$ which enforce smooth, yet flexible, transitions for the group~membership vectors across contiguous age classes, while informing these vectors through external macro--classifications of causes of death. Such a latter exogenous information is encoded in the vector $\bw=(w_1, \ldots, w_n)$, whose generic categorical entry $w_i$ identifies the macro--category $l=1, \ldots, L$ (e.g., neoplasms, malformations,~circulatory~diseases, mental health problems, etc...) of cause $i=1, \ldots, n$. This macro--classification~can~be~retrieved~from the \textsc{icd} system presented in Section~\ref{sec_1.D}, yielding a total of $L=19$ macro--categories.
    \vspace{-2pt}

\setlength\abovedisplayskip{4pt}%
    
Consistent with the above discussion, we leverage the dependent random partition priors  of \citet{page2022dependent} in combination with product partition model constructions   \citep{muller2011product} to obtain
\begin{eqnarray}
\begin{split}
(\bz_{(1)1}, \ldots,  \bz_{(1)m} \mid \bw,  \balpha_{(1)}, \eta_{(1)}) \sim \mbox{\textsc{drpm}-w}(\balpha_{(1)}, \eta_{(1)}),\\
(\bz_{(2)1}, \ldots,  \bz_{(2)m} \mid \bw,  \balpha_{(2)}, \eta_{(2)}) \sim \mbox{\textsc{drpm}-w}(\balpha_{(2)}, \eta_{(2)}),
\label{eq:model:prior_tprm}
\end{split}
\end{eqnarray}
where $\balpha_{(1)}, \balpha_{(2)}, \eta_{(1)}$ and $\eta_{(2)}$  are prior hyper--parameters whose meaning will be clarified in the following, whereas \textsc{drpm}-w denotes the assumed dependent random partition prior informed by $\bw$.~Such a prior is presented in detail in Section~\ref{sec_321} for the generic sequence $\bz_{1}, \ldots,  \bz_{m} $, where the indexes $(1)$ and $(2)$ associated with the two partitions under analysis are removed for the sake of generality.

\subsubsection{Constructive Representation}\label{sec_321}
\vspace{-9pt}
The  $\mbox{\textsc{drpm}-w}(\balpha, \eta)$ prior for a generic sequence  of group allocations $\bz_{1}, \ldots,  \bz_{m} $ is defined hierarchically through a Markovian mechanism, which combines two main  constructions. The first one is a flexible, yet smooth, transition mechanism from $\bz_{x-1}$ to $\bz_{x}$ regulated by $n$ independent $\mbox{Bernoulli}(\alpha_x)$ variables $\gamma_{1x}, \ldots, \gamma_{nx}$, with the generic $\gamma_{ix}$ controlling wether the $i$--th cause of death is allowed to possibly change group allocation from age class $x-1$ to $x$ (i.e., $\gamma_{ix}=0$),~or~not (i.e., $\gamma_{ix}=1$) \citep{page2022dependent}. The second is a carefully--designed supervised Chinese restaurant~process prior $\mbox{\textsc{crp}-w}(\eta)$~\citep[][]{muller2011product,page2022dependent,legramanti2022extended} regulating both the formation of the group structures among causes at the first age class, i.e., $\bz_1$, and the mechanism~through which the subset of causes~with $\gamma_{ix}=0$, for $i=1, \ldots, n$, are re--allocated to clusters~from~$\bz_{x-1}$ to $\bz_{x}$, for each age $x=2, \ldots, m$. 

\setlength\abovedisplayskip{6pt}%
    \setlength\belowdisplayskip{5pt}%
\vspace{-2pt}
Recalling classical results from Bayesian nonparametrics (see \citet{gershman2012tutorial}~for~an~introductory overview) in connection with covariate--dependent product partition models \citep{muller2011product}, the assumed  $\mbox{\textsc{crp}-w}(\eta)$ prior has probability mass function for the generic group membership vector $\bz$ defined as $p(\bz \mid \eta,\bw)\propto[\eta^{H}\Gamma(\eta)/\Gamma(\eta+n)] \prod_{h=1}^{H}\rho_{\textsc{dc}}(\bw_{h},\ba^{\bw})(n_{h}-1)!$.~In~this~expression,~$\Gamma(\cdot)$ corresponds to the Gamma function, $n_{h}$ denotes the number of causes in group~$h$,~and~$\bw_{h}$ is the  vector encoding the memberships to the $L$ external macro--categories of the $n_{h}$ causes~in~group~$h$.~This exogenous information enters $p(\bz \mid \eta,\bw)$ via~the Dirichlet--categorical cohesion function $\rho_{\textsc{dc}}(\cdot, \ba^{\bw})$ with parameters $\ba^{\bw}=(a_1^{\bw}, \ldots, a_L^{\bw})$ \citep[][Ch.\ 4]{muller2011product}, which increases the prior probabilities of those groups that are homogenous with respect to the external macro--classification~of~the~causes.~As such, by including  $\rho_{\textsc{dc}}(\bw_{h},\ba^{\bw})$, the $\mbox{\textsc{crp}-w}(\eta)$~supervises the classical $\mbox{\textsc{crp}}(\eta)$ prior by the external covariate $\bw$ in a way that favors internally--homogenous groups of causes with respect to the associated macro--classification. Such a mechanism is evident when studying the conditional distribution for each $z_{i}$ given $\bz^{(-i)}=(z_{1}, \ldots, z_{i-1},z_{i+1},z_{n})$. Under the $\mbox{\textsc{crp}-w}(\eta)$ prior, this conditional distribution is
\vspace{-2pt}
\begin{equation}\label{eq:model:supervGT_predict_application}
    \mbox{pr}(z_{i} = h \mid \bz^{(-i)},\eta,  \bw) \propto \begin{cases} 
        \displaystyle n^{(-i)}_{h} 
        \frac{\bar{n}^{(-i)}_{h w_{i}}+a^{\bw}_{w_i}}{\bar{n}^{(-i)}_{h\bullet} +  a_\bullet^{\bw}} 
        & \qquad \text{for} \ h = 1, \ldots, H^{(-i)}, \\ 
        \displaystyle \eta
        \frac{a^{\bw}_{w_{i}}}{a_\bullet^{\bw}} 
        & \qquad \text{for} \ h = H^{(-i)} + 1,
    \end{cases} 
\end{equation}
for every $i=1, \ldots, n$, where $H^{(-i)}$ and $n^{(-i)}_{h}$ are the total number of  non--empty groups and the cardinality of the $h$--th~cluster, respectively, after removing cause $i$, while $\bar{n}^{(-i)}_{h w_{i}}$ corresponds to the number of causes in group $h$ with the same macro--category of the $i$--th one. Finally, $\bar{n}^{(-i)}_{h\bullet} = \sum_{l=1}^{L} \bar{n}^{(-i)}_{hl}= |\bw^{(-i)}_h|$, $a_\bullet^{\bw}=\sum_{l=1}^{L} a^{\bw}_l$ and $a^{\bw}_{w_i}$ is the entry of $\ba^{\bw}$ corresponding to the macro--category of cause $i$. According to \eqref{eq:model:supervGT_predict_application}, conditioned on $\bz^{(-i)}$, the $i$--th cause can either occupy a group already observed for the other causes, or generate a new one. The former event has probability which depends on the cardinality~of the group, as for the classical \textsc{crp} prior, further reinforced by a factor that favors
the attribution of the $i$--th cause to those existing groups that have a higher fraction of causes with its same macro--category. The creation of a new cluster is  instead regulated by the $\textsc{crp}$ concentration parameter~$\eta>0$~and~by those of the Dirichlet--categorical cohesion function (i.e., $\ba^{\bw}=(a_1^{\bw}, \ldots, a_L^{\bw})$, with $a_{l}>0$, $l=1, \ldots, L$). Consistent with \eqref{eq:model:supervGT_predict_application}, larger values for $\eta$ yield a higher expected number of clusters in $\bz$.

\setlength\abovedisplayskip{7pt}%
    \setlength\belowdisplayskip{8pt}%
    
Besides illustrating the tractability of the $\mbox{\textsc{crp}-w}(\eta)$ prior, along with the meaning of its parameters, the scheme in \eqref{eq:model:supervGT_predict_application} plays also a key role in the Gibbs sampling algorithm designed in Section~\ref{sec_3}, and clarifies that the total number of groups in the generic  $\bz$ does not need to be pre--specified, but rather can be learned automatically. This is a substantial gain relative to popular \textsc{sbm}s for dynamic~networks that could be possibly applied to our motivating application \citep[see, e.g.,][]{yang2011detecting,xu2014dynamic,xu2015stochastic,matias2017statistical}. In fact, such models assume knowledge~of the total number~of groups, or leverage information criteria which require estimation under different settings for $H$. This is a computationally challenging task in contexts where the number of groups varies with $x$.

Combining the above $\mbox{\textsc{crp}-w}(\eta)$ formulation with the previously--introduced Bernoulli transition mechanism yields the following hierarchical construction for the $\mbox{\textsc{drpm}-w}(\balpha, \eta)$ prior on the generic sequence $\bz_{1}, \ldots,  \bz_{m} $ of group membership vectors
\begin{eqnarray}
\begin{split}
&(\gamma_{ix} \mid \balpha=(\alpha_2, \ldots, \alpha_m)) \stackrel{\mbox{\tiny indep.}}{\sim} \mbox{Bernoulli}(\alpha_x), \qquad i=1, \ldots, n, \ \ x=2, \ldots, m,
\label{eq:model:prior_tprm_hierarch} \\
& (\bz_1 \mid \eta, \bw) \sim \mbox{\textsc{crp}-w}(\eta), \qquad (\bz_x \mid \bz_{x-1}, \bgamma_x, \eta, \bw) \sim \mbox{\textsc{crp}-w}(\eta)|_{\mathcal{P}_{\bz_{x-1}, \bgamma_x}}, \ \ x=2, \ldots, m,
\end{split}
\end{eqnarray}
where $\bgamma_x=(\gamma_{1x},\ldots, \gamma_{nx})$, while $\mbox{\textsc{crp}-w}(\eta)|_{\mathcal{P}_{\bz_{x-1}, \bgamma_x}}$ is the supervised Chinese restaurant process prior constrained to the set of partitions $\mathcal{P}_{\bz_{x-1}, \bgamma_x}$ {\em compatible} with $\bz_{x-1}$ under $\bgamma_x$, namely~all~the~partitions among the $n$ causes of death that can be derived from the one associated with the group~membership vector $\bz_{x-1}$ by reallocating only those causes for which $\gamma_{ix}=0$. Recalling \citet{page2022dependent}, under this constraint, $\mbox{pr}(\bz_x=\bz\mid  \bz_{x-1}, \bgamma_x, \eta, \bw) \propto p(\bz \mid \eta,\bw) \id(\mathbb{Z} \in \mathcal{P}_{\bz_{x-1}, \bgamma_x})$, where $p(\bz \mid \eta,\bw)$ is the previously--defined probability mass function of the $\mbox{\textsc{crp}-w}(\eta)$ prior, whereas $\mathbb{Z}$ is the partition of the $n$ causes associated with the generic group membership vector $\bz$. As such, the prior on $\bz_{x}$~is anchored to $\bz_{x-1}$~by allowing only a subset of causes to change group allocation, with the size of this subset regulated by the parameter $\alpha_x \in [0,1]$.  When $\alpha_x= 1$ the clustering structures among causes~of death at age classes $x-1$ and $x$ perfectly overlap, whereas a value of $\alpha_x= 0$ forces all causes to be re--allocated according to an unrestricted $\mbox{\textsc{crp}-w}(\eta)$ prior, thereby removing the dependence on $\bz_{x-1}$. Therefore, as $\alpha_x$ ranges from $0$ to $1$, the vectors $\bz_{x-1}$ and $\bz_{x}$  are favored~to~be~increasingly~similar. 

Note that the above notion of smoothness  is allowed to change flexibly  across age classes~via~transition--specific parameters $\alpha_2, \ldots, \alpha_m$, which are assigned conjugate Beta hyperpriors, i.e., $\alpha_x \sim \mbox{Beta}(a_{\alpha}, b_{\alpha})$, for $x=2, \ldots, m$. This allows smoothness to be inferred adaptively from the observed matrices~$\bY_1,\ldots, \bY_{m}$. Such a data--oriented perspective is considered also for the \textsc{crp} concentration parameter $\eta$ on which~we assume a conjugate Gamma hyperprior \citep{escobar1995bayesian}, namely $\eta \sim \mbox{Gamma}(a_{\eta},b_{\eta})$. 

As illustrated in Section~\ref{sec_3}, this prior elicitation facilitates the design of a tractable tempered Gibbs--sampler for posterior inference on the sequences of partitions $\bz_{(1)1}, \ldots,  \bz_{(1)m}$ and $\bz_{(2)1}, \ldots,  \bz_{(2)m}$.

\vspace{-9pt}
\section{Posterior Computation and Inference}\label{sec_3}
\vspace{-7pt}
Posterior inference for the sequences of partitions $\bz_{(1)1}, \ldots,  \bz_{(1)m}$ and $\bz_{(2)1}, \ldots,  \bz_{(2)m}$ that parameterize the Bayesian model presented in Section~\ref{sec_2} is performed via Monte Carlo leveraging samples produced by a carefully--designed collapsed Gibbs sampler in combination with adaptive parallel tempering.~This algorithm is derived in detail in Sections \ref{subsec:collapsed_gibbs_sampler}--\ref{subsec:adaptive_parallel_tempering} by exploiting the tractability of the likelihood in  \eqref{eq:model:categ_marg_lik} along with the hierarchical representation \eqref{eq:model:prior_tprm_hierarch} of the priors in \eqref{eq:model:prior_tprm}. Section~\ref{subsec:posterior_summary}, concludes by presenting useful inferential strategies to perform point estimation and uncertainty quantification on the group structures encoded in $\bz_{(1)1}, \ldots,  \bz_{(1)m}$ and $\bz_{(2)1}, \ldots,  \bz_{(2)m}$ leveraging the Gibbs samples.

\vspace{-5pt}
\subsection{Collapsed Gibbs Sampler}%
\vspace{-5pt}
\label{subsec:collapsed_gibbs_sampler}

The proposed Gibbs algorithm samples iteratively from the full--conditional distributions of the group membership vectors $\bz_{(1)1}, \ldots,  \bz_{(1)m}$ and $\bz_{(2)1}, \ldots,  \bz_{(2)m}$, the persistency variables $\bgamma_{(1)2}, \ldots,  \bgamma_{(1)m}$ and $\bgamma_{(2)2}, \ldots,  \bgamma_{(2)m}$ appearing in the hierarchical representation \eqref{eq:model:prior_tprm_hierarch} of the \textsc{drpm}-w priors in \eqref{eq:model:prior_tprm}, and finally, the hyperparameters $\balpha_{(1)}$, $\eta_{(1)}$, $\balpha_{(2)}$, and $\eta_{(2)}$. As clarified in the following, by combining the priors presented in Section~\ref{sec_32} with the collapsed likelihood in  \eqref{eq:model:categ_marg_lik} yields  closed--form full--conditionals for all these quantities, thereby facilitating the design of a tractable sampling scheme.

\setlength\abovedisplayskip{5pt}%
    \setlength\belowdisplayskip{6pt}%
    
Focusing on the row--specific quantities $\bz_{(1)1}, \ldots,  \bz_{(1)m}$, $\bgamma_{(1)2}, \ldots,  \bgamma_{(1)m}$, $\balpha_{(1)}$ and $\eta_{(1)}$,~let~us~first~introduce some useful notation. In particular, let $\Gamma_{(1)x}=\{i=1, \ldots, n: \gamma_{(1)ix}=1 \}$~be~the~set~of~underlying causes whose group allocation does not change from $x-1$ to $x$. Moreover, denote with $\mathbb{Z}_{(1)x|\Gamma_{(1)x}}$ the partition induced by $\bz_{(1)x}$ considering only those underlying causes with indexes in $\Gamma_{(1)x}$. Then, adapting the derivations in \citet{page2022dependent} to our construction, the full--conditional distributions~for the binary persistency variables $\gamma_{(1)ix}$, $i=1, \ldots, n$, $x=2, \ldots, m$, are Bernoulli with probabilities
\begin{equation}
    \pr(\gamma_{(1)ix} = 1 \mid -) = 
    \frac{\alpha_{(1)x}}{\alpha_{(1)x} + (1 - \alpha_{(1)x}) \, {\displaystyle\frac{p(\mathbb{Z}_{(1)x|\Gamma^{(+i)}_{(1)x}} \mid \eta_{(1)}, \bw)}{p(\mathbb{Z}_{(1)x|\Gamma^{(-i)}_{(1)x}} \mid \eta_{(1)}, \bw )}}} \id[\mathbb{Z}_{(1)x|\Gamma^{(+i)}_{(1)x}}=\mathbb{Z}_{(1)x-1|\Gamma^{(+i)}_{(1)x}}],
    \label{full-cond-gamma}
\end{equation}
 for  each node $i=1, \ldots, n$ and age class $x=2, \ldots, m$, where $\id[\cdot]$ corresponds to the indicator function, whereas $\Gamma^{(-i)}_{(1)x}=\Gamma_{(1)x} \setminus \{i\}$ and $\Gamma^{(+i)}_{(1)x}=\Gamma^{(-i)}_{(1)x} \cup \{i\}$. Notice that within the above expression the~ratio $p(\mathbb{Z}_{(1)x|\Gamma^{(+i)}_{(1)x}} \mid \eta_{(1)}, \bw)/p(\mathbb{Z}_{(1)x|\Gamma^{(-i)}_{(1)x}} \mid \eta_{(1)}, \bw )$ relates directly to the conditional distribution for $z_{(1)ix}$ given $\bz_{(1)x}^{(-i)}$, $ \eta_{(1)}$ and $\bw$. As such, recalling also \citet{page2022dependent},~it~can~be~computed~from~\eqref{eq:model:supervGT_predict_application} after replacing the generic quantities $n^{(-i)}_h$, $\bar{n}^{(-i)}_{h w_i}$,  $\bar{n}^{(-i)}_{h \bullet}$ and $\eta$ with those specific to the reduced partition~among~the underlying causes at age class $x$.

The results in \eqref{eq:model:supervGT_predict_application} are also useful to sample the allocations $z_{(1)ix}$ at each age class $x=1, \ldots, m$ for those underlying causes whose sampled $\gamma_{(1)ix}$ is equal to $0$; if $\gamma_{(1)ix}=1$, then $z_{(1)ix}$ is kept fixed to the group allocation drawn for cause $i$ at $x-1$. More specifically, denote with $\mathbb{Z}_{(1)x}^{z_{(1)ix}=h}$~the~partition~induced by the group membership vector $(z_{(1)1x}, \ldots,z_{(1)ix}=h ,\ldots z_{(1)nx})$ with the $i$--th underlying cause allocated to cluster $h$. Then, direct application of the Bayes rule in combination with results~from~\citet{page2022dependent} yields a full--conditional categorical distribution for each $z_{(1)ix}$ with probabilities 
\begin{equation}\label{eq:inference:fullcond_cluster}
\begin{split}
    &\pr(z_{(1)ix} = h \mid -)\\
    & \ \ \ \ \  \propto \pr(z_{(1)ix} = h \mid \bz_{(1)x}^{(-i)},\eta_{(1)},\bw)\id[\mathbb{Z}_{(1)x|\Gamma_{(1)x+1}}^{z_{(1)ix}=h}=\mathbb{Z}_{(1)x+1|\Gamma_{(1)x+1}} ]\frac{p(\bY_x \mid z_{(1)ix} = h, \bz_{(1)x}^{(-i)}, \bz_{(2)x})}{p(\bY_{x}^{(-i\cdot)} \mid \bz_{(1)x}^{(-i)}, \bz_{(2)x})} ,
    \end{split}
\end{equation}
for every $h=1, \ldots, H^{(-i)}_{x}+1$, $i=1, \ldots, n$ and $x=1, \ldots, m$, where $\bY_{x}^{(-i\cdot)}$ corresponds to~the~adjacency~matrix at~age class $x$ without the $i$--th row. In the above expression, the~prior~probabilities $\pr(z_{(1)ix} = h \mid \bz_{(1)x}^{(-i)},\eta_{(1)},\bw)$ are available directly from \eqref{eq:model:supervGT_predict_application} after replacing the involved generic quantities with those specific to the partition~of~the underlying causes at age $x$, whereas~the constraint included through the indicator function guarantees the compatibility discussed in Section~\ref{sec_321} for contiguous partitions (notice that compatibility with the partition at age class $x-1$~holds by prior construction, and hence, it does not need to be checked). Finally, generalizing results on Beta--Binomial distributions \citep[e.g.,][]{schmidt_2013,legramanti2022extended} to the Dirichlet--categorical one in~\eqref{eq:model:categ_marg_lik}, also the likelihood factor within \eqref{eq:inference:fullcond_cluster} admits the closed--form expression
\begin{equation}
   \frac{p(\bY_x \mid z_{(1)ix} = h, \bz_{(1)x}^{(-i)}, \bz_{(2)x})}{p(\bY_{x}^{(-i\cdot)} \mid \bz_{(1)x}^{(-i)}, \bz_{(2)x})} = \prod_{k = 1}^{K_x} \frac{\Gamma(a^\theta_{\bullet} + n_{hkx\bullet}^{(-i)})}{\Gamma(a^\theta_{\bullet} + n_{hkx\bullet})} \prod_{c = 1}^4 \frac{\Gamma(a^\theta_{c} + n_{hkxc})}{\Gamma(a^\theta_{c} + n^{(-i)}_{hkxc})},
    \label{lik_ratio_under}
\end{equation}
for every $h=1, \ldots, H^{(-i)}_{x}+1$, $i=1, \ldots, n$ and $x=1, \ldots, m$. 

\setlength\abovedisplayskip{7pt}%
    \setlength\belowdisplayskip{8pt}%
    
Given the persistency variables $\gamma_{(1)ix}$, $i=1, \ldots, n$, $x=2, \ldots, m$, also the prior hyperparameters $\alpha_{(1)x}$, $x=2, \ldots, m$ admit closed--form full conditional distributions. More specifically, combining the Bernoulli likelihood for each $\gamma_{(1)ix}$ with the Beta priors assumed~in~Section~\ref{sec_321} for every $\alpha_{(1)x}$,  yields
\begin{eqnarray}
    (\alpha_{(1)x} \mid -) \sim \dBeta \left(a_\alpha + \sum\nolimits_{i=1}^n \gamma_{(1)ix}, b_\alpha + n - \sum\nolimits_{i = 1}^n \gamma_{(1)ix} \right), 
    \label{eq:inference:fullcond_beta}
\end{eqnarray}
independently for $x=2, \ldots, m$. Similarly--tractable full--conditionals are also available~for~the~\textsc{crp}~concentration parameter $\eta_{(1)}$ via a direct application of the results in \citet{escobar1995bayesian}.

To conclude, it remains to update the column--specific~quantities~$\bz_{(2)1}, \ldots,  \bz_{(2)m}$, $\bgamma_{(2)2}, \ldots,  \bgamma_{(2)m}$,~$\balpha_{(2)}$ and $\eta_{(2)}$. Due to the model symmetry, these updates share the same structure~of those derived for the row--specific counterparts. As a consequence, it suffices to apply again~\eqref{full-cond-gamma}--\eqref{eq:inference:fullcond_beta} by replacing the quantities and indexes related to the underlying causes with those~of~the contributing~ones,~and~vice--versa.

\vspace{-7pt}
\subsection{Adaptive Parallel Tempering}%
\label{subsec:adaptive_parallel_tempering}
\vspace{-6pt}
Albeit tractable, the Gibbs routine presented in Section~\ref{subsec:collapsed_gibbs_sampler}  requires exploration of a high--dimensional discrete space involving two sequences of dependent random partitions. To mitigate the possible~mixing issues that may arise in this challenging computational setting, we combine the previously--derived Gibbs sampler with an adaptive parallel tempering implementation. 

Parallel tempering (\textsc{pt}) is a general methodology to improve the mixing of \textsc{mcmc} algorithms,~especially in settings characterized by complex high--dimensional posteriors that may exhibit multiple local modes \citep[e.g.,][]{earl&deem2005, syed&al2022}.
In its general form, \textsc{pt} runs a sequence of independent \textsc{mcmc} samplers at different \emph{temperatures}, where higher--temperature chains allow for~an improved exploration of the state space by flattening the posterior distribution, and hence,~also~its~local modes.  In contrast, lower--temperature chains sample from distributions progressively similar~to~the~actual posterior, thereby targeting the original distribution of interest. At each iteration, all chains are independently updated performing a local move, which, in our case, uses the Gibbs sampling steps outlined in Section~\ref{subsec:collapsed_gibbs_sampler}. Then, at periodic intervals, the algorithm attempts to swap states between chains~via~a Metropolis--Hastings acceptance criterion that ensures detailed--balance is maintained.  This swap allows lower--temperature chains to benefit from the broader exploration of higher--temperature~ones.

Within our implementation we consider, in particular, non--reversible deterministic swap proposals \citep{syed&al2021} to guarantee a more rapid information exchange among high-- and low--temperature chains.
Moreover, we implement the online stochastic optimization method proposed by \cite{miasojedow&at2013} to dynamically adapt the temperature grid during the \textsc{pt} runs.
This~strategy~builds a decreasing sequence of inverse temperatures and, at each swap, it uses the associated acceptance~probabilities to tune the inverse--temperature grid via a Robbins--Monroe update targeting the desired acceptance level.
Such a scheme robustifies  \textsc{pt} against suboptimal specifications~of~the~initial~temperature schedule, thus allowing for a more efficient exchange of information from high-- to low--temperature chains, which is fundamental to obtain an effective exploration of the posterior~distribution.

\vspace{-5pt}

\subsection{Posterior Inference}%
\label{subsec:posterior_summary}
\vspace{-7pt}

Posterior inference on the quantities of interest is performed via Monte Carlo, relying on the samples produced by the routine outlined in Sections~\ref{subsec:collapsed_gibbs_sampler}--\ref{subsec:adaptive_parallel_tempering}. As discussed in Sections~\ref{sec_1}--\ref{sec_2}, our primary focus lies on the group membership structures encoded in the sequences $\bz_{(1)1}, \ldots,  \bz_{(1)m}$ and $\bz_{(2)1}, \ldots,  \bz_{(2)m}$, which unveil the block patterns displayed by the observed~causes--of--death networks across age classes. Extending general guidelines in Bayesian model--based clustering \citep[e.g.,][]{wade2018} to our specific context, a natural and interpretable option for summarizing the posterior distribution over these group structures is to compute the so--called posterior  {\em similarity} (or {\em co--clustering}) matrices $\hat\bP_{(1)1}, \ldots, \hat\bP_{(1)m}$ and $\hat\bP_{(2)1}, \ldots, \hat\bP_{(2)m}$, whose generic entries $\hat\bP_{(1)x[i,i']}$ and $\hat\bP_{(2)x[j,j']}$ estimate $\pr(z_{(1)ix} = z_{(1)i'x} \mid \bY_1, \ldots, \bY_m)$ and $\pr(z_{(2)jx} = z_{(2)j'x} \mid \bY_1, \ldots, \bY_m)$ via the relative frequency of  Gibbs samples in which $z_{(1)ix}=z_{(1)i'x}$ and  $z_{(2)jx}=z_{(2)j'x}$, respectively. This yields an interpretable probabilistic summary of the co--clustering relationships supported by the posterior across age classes, along with~a quantification of uncertainty in the inferred group structures.

\setlength\abovedisplayskip{5pt}%
    \setlength\belowdisplayskip{6pt}%
    
    \vspace{-2pt}
Leveraging the above {\em similarity} matrices in combination with the decision--theoretic framework of \citet{wade2018} it is also possibile to obtain the posterior point estimates $\hat\bz_{(1)x}$ and~$\hat\bz_{(2)x}$ for $\bz_{(1)x}$ and $\bz_{(2)x}$, respectively, at each age class $x=1, \ldots, m$, via
\begin{eqnarray*}
    \hat\bz_{(1)x} = \mathop{\mbox{argmin}}\nolimits_{\bz} \mathbb{E}[ \,\textsc{vi}(\bz_{(1)x}, \bz) \mid \bY_1, \ldots, \bY_m], \ \ \ \mbox{and} \  \ \ \hat\bz_{(2)x} = \mathop{\mbox{argmin}}\nolimits_{\bz} \mathbb{E}[ \,\textsc{vi}(\bz_{(2)x}, \bz) \mid \bY_1, \ldots, \bY_m],
\end{eqnarray*}
where  $\textsc{vi}(\cdot,\cdot)$ is the variation of information distance introduced by \cite{meilua2007comparing} as a metric to measure differences between two generic partitions via a comparison among individual and joint entropies. The solutions to the above minimization problems can be obtained under the \verb|R| library \verb|mcclust.ext|  \citep{wade2018}, which requires, as inputs, the matrices $\hat\bP_{(1)x}$ and $\hat\bP_{(2)x}$, respectively. 
    \vspace{-2pt}

Besides studying the age--specific group structures among underlying and contributing causes encoded within $\hat\bz_{(1)x}$ and~$\hat\bz_{(2)x}$, respectively, in our specific context it is also of interest to assess the stability of these structures across age classes. To this end, in Section~\ref{sec_5} we will further~study~the~\emph{meet}~of the estimated partitions across selected contiguous ages. Recalling, e.g., \citet{wade2018}, such a meet corresponds to a finer partition where two causes belong to the same \emph{meet cluster}~if~these causes co--cluster in all the considered partitions, thereby highlighting sets of underlying and contributing causes that display stable group behaviors across the selected age classes.
    \vspace{-2pt}

\setlength\abovedisplayskip{3pt}%
    \setlength\belowdisplayskip{3pt}%
To conclude, notice that while the block--specific vectors $\btheta_{hkx}$ are integrated out to obtain  the~likelihood~in~\eqref{eq:model:categ_marg_lik}, when such quantities are also of interest a plug--in estimate  can be readily derived.~In~particular, denote with $\hat{n}_{hkxc}$ the total number of pairs $(i,j)$ such that $\hat{z}_{(1)ix}=h$, $\hat{z}_{(2)jx}=k$~and~$y_{ijx}=c$. Then, leveraging Dirichlet--categorical conjugacy, a sensible point estimate for each $\theta_{hkxc}$ is
\begin{equation}
    \hat\theta_{hkxc} 
    = \mathbb{E}(\theta_{hkxc} \mid \bY_1, \ldots, \bY_m, \hat\bz_{(1)x}, \hat\bz_{(2)x}) 
    = (a^\theta_{c} + \hat n_{hkxc})/(a^\theta_{\bullet} + \hat n_{hkx\bullet}), 
    \label{teta_plug_est}
\end{equation}
for every $h =  1, \dots, \hat{H}_x $, $k = 1, \dots, \hat{K}_x $, $x=1, \ldots, m$ and $c =1, \ldots, 4$.

\setlength\abovedisplayskip{7pt}%
    \setlength\belowdisplayskip{8pt}%
    
\section{Simulation Studies}\label{sec_4}
\vspace{-12pt}
In this section, we present extensive simulation studies that illustrate the performance of the model proposed in Section~\ref{sec_2} and quantify its gains with the respect to state--of--the--art alternatives that could be possibly employed within our motivating application.  As a benchmark competitor, we consider, in particular, the dynamic \textsc{sbm} proposed by \citet{matias2017statistical} and implemented~in~the~\verb|R|~library \verb|dynsbm|. This formulation has emerged in recent years as one of the most widely adopted implementations for inference on group structures among nodes varying across a temporal index \citep[][]{peixoto2018, kim&al2018,lee2019review}, making it a natural candidate for benchmarking~in our simulation study. Note that, unlike for our proposed~model, the one designed by \citet{matias2017statistical} does not allow for two separate partitions on the rows and columns, respectively, and further assumes that the total number of non--empty groups is constant across the sequence of networks.~As~such, the comparison against this competitor is also useful to assess whether the increased flexibility provided~by the model we propose yields practical gains in settings aligned with our motivating application.

    \vspace{-2pt}

\begin{figure}[t!]
    \centering
    \includegraphics[width=0.95\linewidth]{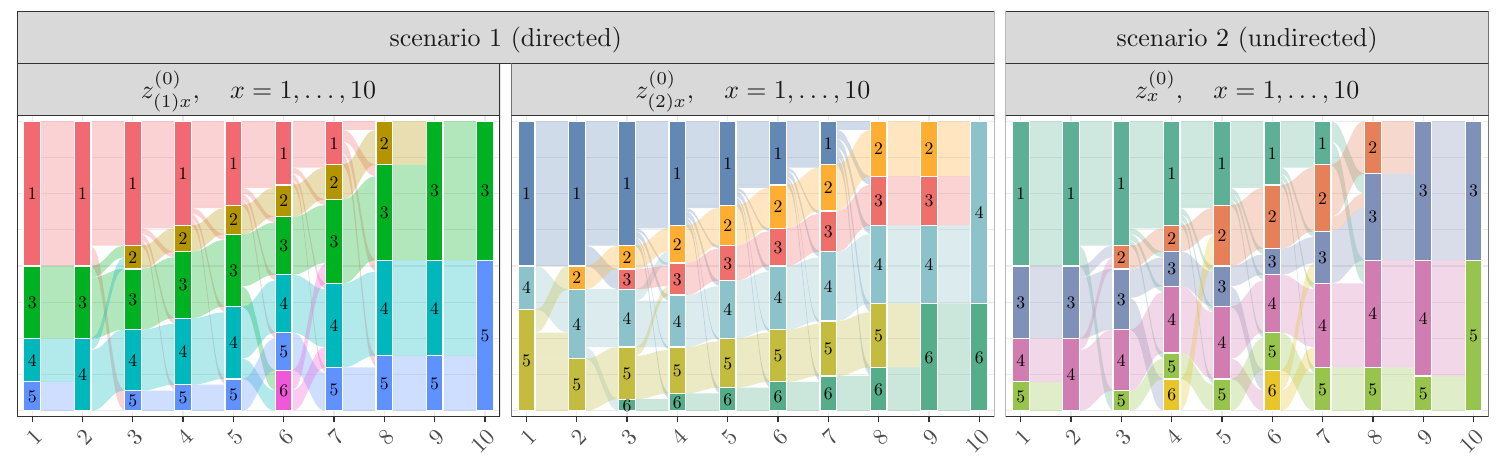}
    \vspace{-11pt}
    \caption{\footnotesize Riverplots representing the evolution of $\bz^{(0)}_{(1)x}$ and $\bz^{(0)}_{(2)x}$, $x=1, \ldots, 10$ in the first scenario, and $\bz^{(0)}_{x}$, $x=1, \ldots, 10$ in the second scenario.}
    \label{fig:simstudy:flow}
\end{figure}

\begin{figure}[t!]
    \centering
    \includegraphics[width=1\linewidth]{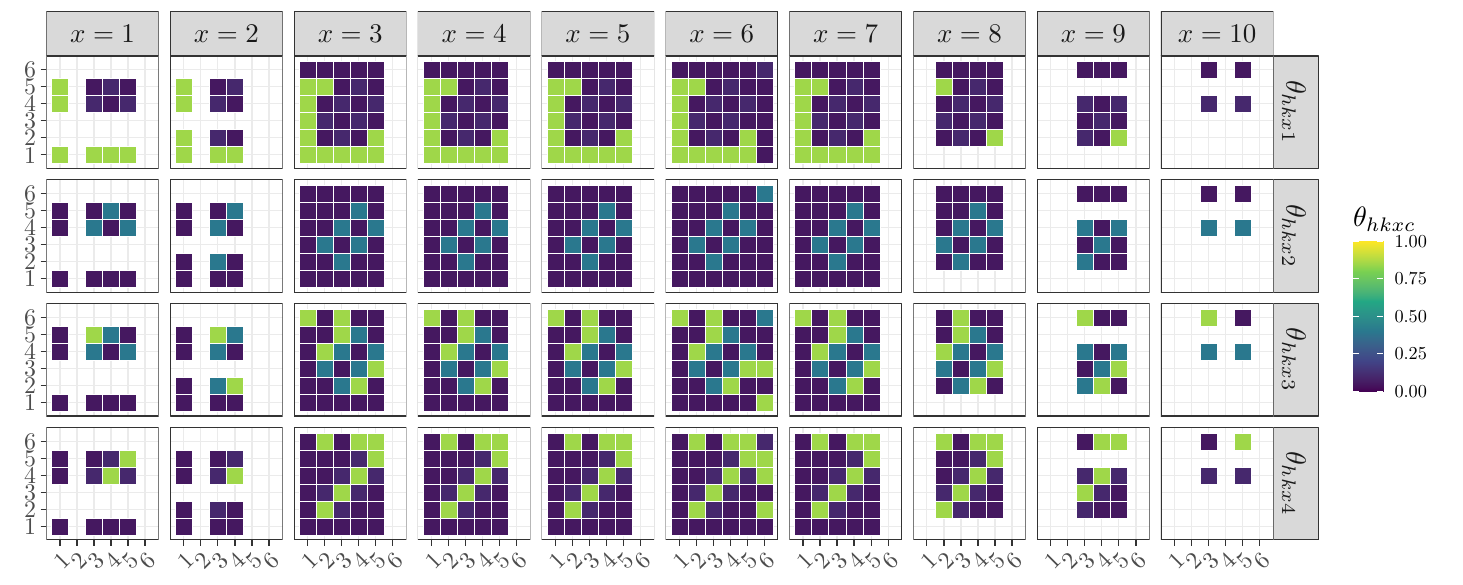}
        \vspace{-20pt}
    \caption{\footnotesize For the first simulation scenario (directed), graphical representation of the block--specific parameters, $\theta^{(0)}_{hkxc}$, evolving over age classes $x=1, \ldots, 10$ and stratified by category $c=1, \ldots, 4$. The numbered rows and columns~correspond to row-- and column--specific groups. White cells correspond to pairs of empty groups at the specific ages.}
    \label{fig:simstudy:theta}
\end{figure}

To provide a comprehensive assessment, the above models are tested in two different simulation scenarios. The first resembles the structures discussed in Section~\ref{sec_1} for the motivating causes--of--death networks,  where the data are generated from a sequence of categorically--weighted directed networks regulated by two distinct partitions on the rows and columns of the adjacency matrices. As such,~we simulate edges $  (y_{ijx} \mid z^{(0)}_{(1)ix} = h, z^{(0)}_{(2)jx} = k, \btheta^{(0)}_{hkx}) \sim \dCat_{1:4}(\btheta^{(0)}_{hkx})$, independently for~every~$i=1, \ldots, 100$, $j=1, \ldots, 100$ and $x=1, \ldots, 10$, where each $\btheta^{(0)}_{hkx}$ is selected from four possible configurations --- namely, $[0.85, 0.05, 0.05, 0.05]$, $[0.05, 0.05, 0.85, 0.05]$, $[0.05, 0.05, 0.05, 0.85]$ and $[0.1, 0.4, 0.4, 0.1]$ --- (see Figure~\ref{fig:simstudy:theta}), whereas $\bz^{(0)}_{(1)x}$ and $\bz^{(0)}_{(2)x}$ are obtained by manually fixing an evolution across ages for the total of non--empty clusters and the number of causes allowed to change group among consecutive~ages.~Conversely, the selection of which specific causes are considered for these transitions and the updated~group allocations for such causes are performed randomly  (see Figure~\ref{fig:simstudy:flow} for a graphical representation of $\bz^{(0)}_{(1)x}$ and $\bz^{(0)}_{(2)x}$, $x=1, \ldots, 10$). In this way, the partitions to~be inferred are not simulated from the assumed prior, thereby providing a more realistic assessment of the model proposed. For this same~reason,~we further test our construction in a second simulation scenario~consisting of sequences of undirected networks with block structures regulated by a single partition $\bz^{(0)}_{x}$, for $x=1, \ldots, 10$, that is common~to both rows and columns. Recalling  our previous discussion, this setting is favorable to the model of \citet{matias2017statistical} and can be simulated similarly to the directed case by forcing $\bz^{(0)}_{(1)x}=\bz^{(0)}_{(2)x}=\bz^{(0)}_{x}$, $\btheta^{(0)}_{hkx}=\btheta^{(0)}_{khx}$, and $y_{ijx}=y_{jix}$ for each $x=1, \ldots, 10$ (see Figure~\ref{fig:simstudy:flow} for a graphical illustration of  $\bz^{(0)}_{x}$, $x=1, \ldots, 10$).  Notice that, in both scenarios, we do not consider supervision by external covariates~$\bw$ to~avoid~an overly--penalized treatment of the model by  \citet{matias2017statistical}, which is not designed to include this additional source of information. When $\bw$ is not available, it suffices to remove  $(\bar{n}^{(-i)}_{h w_{i}}+a^{\bw}_{w_i})/(\bar{n}^{(-i)}_{h\bullet} +  a_\bullet^{\bw})$ and $a^{\bw}_{w_i}/ a_\bullet^{\bw}$ in \eqref{eq:model:supervGT_predict_application} for adapting our formulation to the unsupervised case.


Given the above simulated data, we perform posterior inference under the Bayesian model proposed within  Section~\ref{sec_2} leveraging diffuse  $\mbox{Dirichlet}(1,1,1,1)$, $\mbox{Beta}(1,1)$ and $\mbox{Gamma}(0.002, 0.001)$ priors~on the block--specific parameters vectors, the transition probabilities and the \textsc{crp} concentration parameters, respectively. Under such settings, posterior samples for $\bz_{(1)1},\ldots, \bz_{(1)10}$ and $\bz_{(2)1},\ldots, \bz_{(2)10}$ in the first simulation scenario can be obtained via the tempered Gibbs--sampler developed in Sections~\ref{subsec:collapsed_gibbs_sampler}--\ref{subsec:adaptive_parallel_tempering}. Such a routine can be also adapted to the second scenario by sampling from the posterior~of~a~single sequence $\bz_{1},\ldots, \bz_{10}$ (having the same prior as $\bz_{(1)1},\ldots, \bz_{(1)10}$ and $\bz_{(2)1},\ldots, \bz_{(2)10}$), while replacing the likelihood induced by directed networks with the one arising in undirected settings, where $y_{ijx}=y_{jix}$, and hence, $\btheta_{hkx}=\btheta_{khx}$. As a consequence, in the directed case, $p(\bY_1, \ldots, \bY_{m} \mid  \bz_{x}, \ x=1, \ldots, m)   
    = \prod_{x = 1}^{m}(\prod_{h = 1}^{H_x} \prod_{k = 1}^{h} [\{\Gamma(a_{\bullet}^\theta)/\Gamma(a_{\bullet}^\theta + n_{hkx\bullet})\} \prod_{c = 1}^{4} \{\Gamma(a_c^\theta + n_{hkxc})/\Gamma(a_c^\theta)\} ])$. 
 
 \vspace{-1pt}
    
 Under the above routines, we consider 4{,}000 posterior samples for the sequences of partitions analyzed. Such samples are obtained after a conservative burn--in of 10{,}000 and thinning by~10~the~subsequent 40{,}000 draws. The study of the traceplots for the logarithm of the likelihoods associated~with the two simulation scenarios does not provide evidence against convergence, and showcase adequate mixing. As such, the produced samples are leveraged to obtain a posterior~point estimate for $\bz_{(1)1},\ldots, \bz_{(1)10}$~and $\bz_{(2)1},\ldots, \bz_{(2)10}$ in the first scenario, and $\bz_{1},\ldots, \bz_{10}$ in the second, via the strategies in Section~\ref{subsec:posterior_summary}.

 \vspace{-1pt}

\begin{figure}[t]
    \centering
    \includegraphics[width=0.92\linewidth]{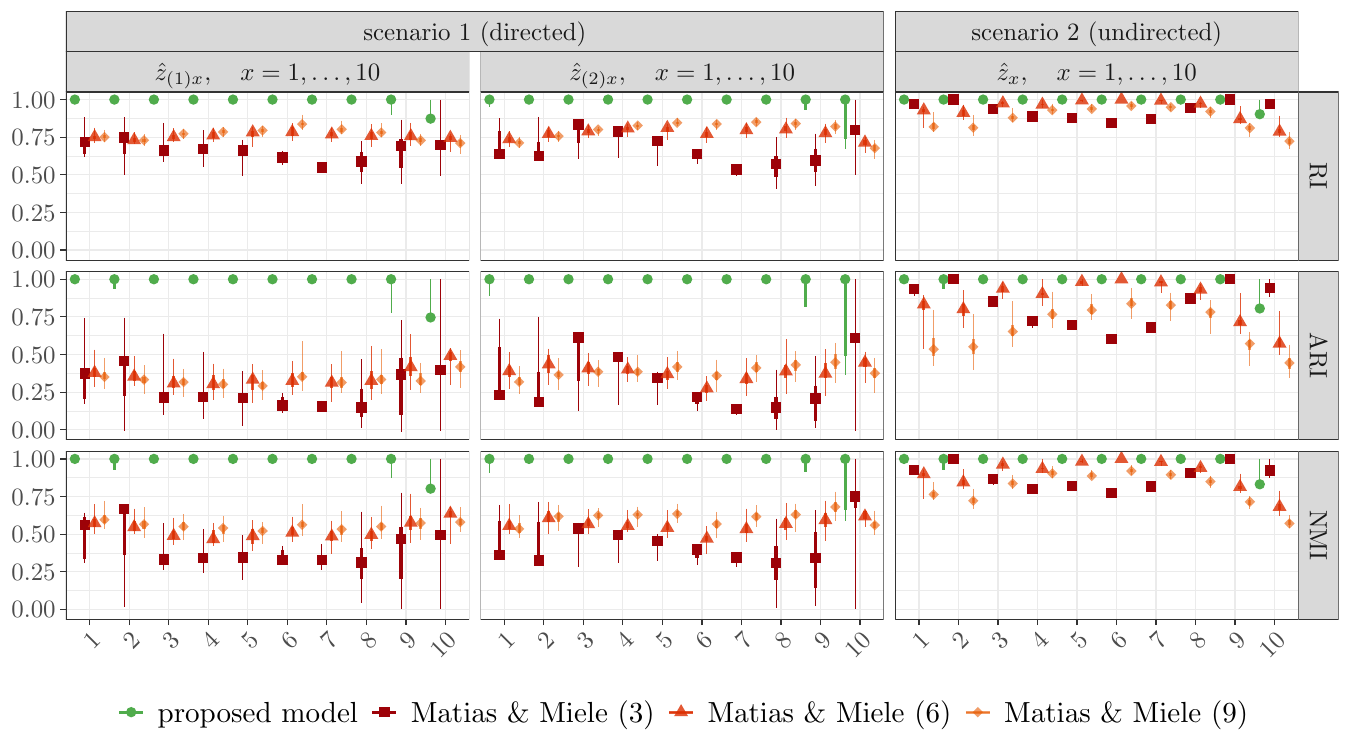}
    \vspace{-10pt}
    \caption{\footnotesize Accuracy of the proposed model and of the one designed by \citet{matias2017statistical} (under three specifications for the total number of clusters) in estimating the true group membership vectors \smash{$\bz^{(0)}_{(1)x}$} and \smash{$\bz^{(0)}_{(2)x}$}, $x=1, \ldots, 10$~in~the first scenario, and \smash{$\bz^{(0)}_{x}$}, $x=1, \ldots, 10$ in the second scenario. The performance is measured under the rand index  (\textsc{ri}), adjusted rand index (\textsc{ari}), and normalized mutual information  (\textsc{nmi}). The points, thick lines, and thin lines represent, respectively, the medians, the inter--quartile ranges, and the ranges of the considered measures~over~50~replicated~experiments of the two simulation scenarios under analysis.}
    \label{fig:simstudy:res}
\end{figure}

Figure~\ref{fig:simstudy:res} illustrates the quality of these estimated partitions in recovering the true ones behind the generative mechanism of the simulated data in scenario one (i.e., $\bz^{(0)}_{(1)1},\ldots, \bz^{(0)}_{(1)10}$ and $\bz^{(0)}_{(2)1},\ldots, \bz^{(0)}_{(2)10}$)  and two (i.e., $\bz^{(0)}_{1},\ldots, \bz^{(0)}_{10}$). Such a quality is quantified in 50 replicated experiments under both~scenarios via the most widely--implemented measures of clustering accuracy, namely,   the rand index (\textsc{ri}) \citep[][]{rand1971}, the adjusted rand index (\textsc{ari})  \citep[][]{hubert&arabie1985}, and the normalized mutual~information (\textsc{nmi}) \citep[][]{strehl2002cluster}. These three measures take values~below~$1$~(which~corresponds to perfect overlap among the two partitions compared), and are computed also for the estimates produced by the model of \citet{matias2017statistical} under its implementation in the  \verb|R| library \verb|dynsbm|. As discussed previously, this competitor does not automatically infer the number of non--empty clusters. As such, we explore three specifications with this quantity set equal to 3, 6 and 9, where~3~corresponds to an underparameterized model, 6 is the true maximum number of clusters in the data, and 9 represents a flexible overparameterized formulation.
Moreover, as suggested by the authors, we obtain~the maximum likelihood estimate of the partitions via a multi--start routine~from~25~different~initial~points.

As shown in Figure~\ref{fig:simstudy:res}, our proposed model almost perfectly estimates all the partitions across ages, in both scenarios, and systematically outperforms the clustering accuracy achieved by the state--of--the--art alternative in \citet{matias2017statistical}. The remarkable gains obtained in the first scenario~clarify the importance of allowing for two separate partitions on the rows and columns in directed networks displaying asymmetric block patterns within the adjacency matrix. As discussed in~Section~\ref{sec_1.D} (see~also Figure~\ref{figure:1}), this is the case of our causes--of--death networks, thereby highlighting~the~need~for a model as the one we propose in Section~\ref{sec_2} to avoid   the substantial bias that would~arise~when~adapting~available solutions to our motivating application. Interestingly, our model outperforms \citet{matias2017statistical} also in the second scenario, where the focus is on those undirected settings favorable to the competitor under analysis. This  clarifies that, besides the increased flexibility provided by the inclusion of two separate partitions in directed settings, the evolution process induced by the prior in Section~\ref{sec_32} on the sequence of partitions appears to be a superior construction in more general settings, including in undirected ones. Unlike for \citet{matias2017statistical}, the process we employ allows the number of non--empty groups to change with ages and to be automatically inferred. These advantages motivate~the use of the proposed model in the analysis of the causes--of--death networks introduced in Section~\ref{sec_1}. A detailed presentation of the important findings obtained from~this analysis is provided in Section~\ref{sec_5}.

\vspace{-9pt}

\section{Application to USA Cause--of--Death Networks}\label{sec_5}
\vspace{-7pt}

We conclude by  illustrating the potential of the proposed model in unveiling group structures and modular interactions among causes of death that were hidden~to~previous studies of the \textsc{usa}~data~presented in Section~\ref{sec_1.D}. In line with standard practice~in~demography, we apply the model separately~to male and female populations. Posterior inference relies on the same hyperparameters and \textsc{mcmc} settings~as in the simulation studies~to~assess the robustness of these default choices in general contexts.

\vspace{-6pt}

\subsection{Empirical Results and Findings}
\begin{figure}[t!]
    \centering
    \includegraphics[width = .95\textwidth]{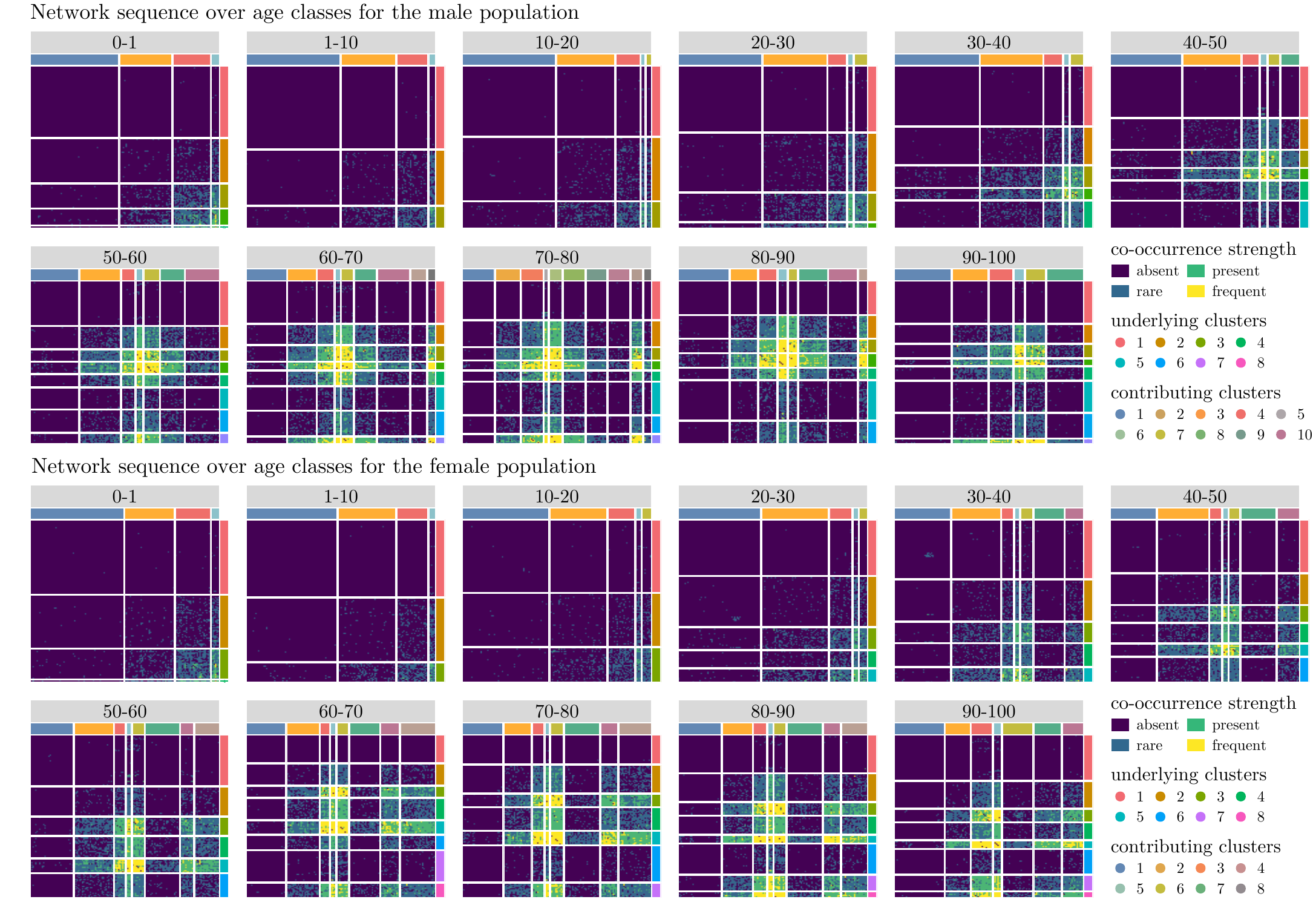}
    \caption{\footnotesize Observed co--occurrence networks among underlying and contributing causes of death at different age classes in the USA male (top) and female (bottom) populations. Rows and columns of the adjacency matrices are re--ordered according to the group structures estimated under the dependent stochastic block model proposed in Section~\ref{sec_2}. } 
    \label{fig:intro:data_plot}
\end{figure}

\vspace{-5pt}
As a first important assessment, Figure~\ref{fig:intro:data_plot} displays the age--specific adjacency matrices of the observed co--occurrence networks among underlying and contributing causes, with rows and columns re--ordered according to the groups estimated by our model. This representation clarifies~that~the~inclusion of two separate partitions varying with age classes is essential to obtain an accurate characterization of the block structures in the observed networks for both males and females. In addition, it confirms that our model provides an highly--effective construction in achieving such an objective, thus motivating~an in--depth analysis of the composition and evolution across ages of the estimated~partitions.

\begin{figure}[t]
    \centering
    \includegraphics[width=0.84\linewidth]{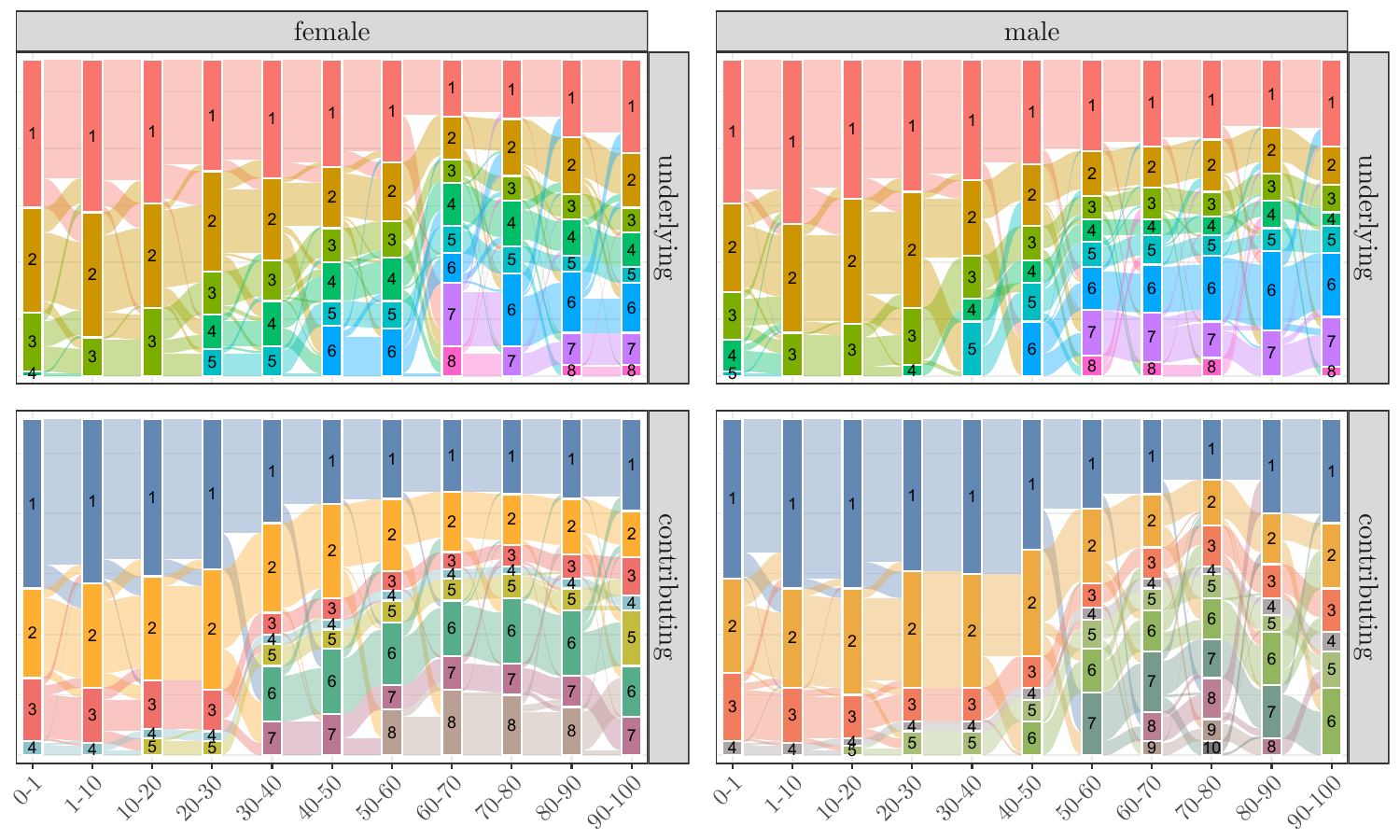}
    \vspace{-5pt}
    \caption{\footnotesize Riverplots representing the evolution of the estimated partitions $\hat \bz_{(1)x}$ and $\hat \bz_{(2)x}$ over age classes $x=1, \ldots, 11$ in the female and male populations. The size of each bar is proportional to the cardinality of the inferred group.}
    \label{fig:application:flow}
\end{figure}

Consistent with the above discussion, Figure~\ref{fig:application:flow} presents the minimum--\textsc{vi} point estimates~of~the~partitions for the underlying and contributing causes across age classes in both the female~and~male~populations. 
The slowly--varying number of groups across different life stages --- ranging from~3~to~8~for females, and from 3 to 10 for males --- further supports the need for a model that allows for smoothly--evolving partition structures across ages.  Notice how the total number of non--empty groups increases progressively with age classes, particularly in the middle and late adulthood, highlighting the growing complexity of the causes--of--death landscape, which reflects the broadening range of health conditions contributing to mortality.  Correspondingly, the average~group~size declines with age classes, thereby unveiling the increasing heterogeneity in the interactions between underlying and contributing causes (see also  Figure~\ref{fig:intro:data_plot}). As is evident from  Figure~\ref{fig:application:flow}, such an heterogeneity results from smooth fragmentations of medium--to--large underlying and contributing groups driven by diverging interaction patterns between specific subsets of causes. These~peculiar~transition~patterns~are~studied~in~detail~in~the~following through the analysis of specific meet clusters, and crucially expand available~findings~on~underlying \citep{bergeron2020diversification,calazans2023levels} and multiple  \citep{trias2023cause} causes--of--death diversity by clarifying that systematic heterogeneity increments are found also in higher--level block--interactions among underlying and contributing causes~over~ages.

Among the aforementioned transition and fragmentation patterns, a relevant one, which has received limited attention in the literature, refers to evolution of group 1 displayed in Figures \ref{fig:intro:data_plot}--\ref{fig:application:flow}~for~both the underlying and contributing causes. As illustrated within Figure~\ref{fig:intro:data_plot}, such a group comprises non--interacting causes which are often removed from the analysis when the focus is on a single age class~or on an aggregated network for a subset of ages. Albeit common, this practice neglects the fact that causes appearing as non--interacting at given age classes may display meaningful co--occurrences~at~other ages, thereby hindering the possibility to infer such peculiar transition mechanism from non--interacting to interacting blocks across age classes. In fact, addressing this objective is a key to understand at which ages specific causes join or exit inactive groups, a crucial information in the design of targeted policies. Consistent with this aim, we retain all the causes of death in our analysis, and let the proposed model infer non--interacting groups along with the associated evolution across ages.  

\vspace{-2pt}
As illustrated in Figures~\ref{fig:intro:data_plot}--\ref{fig:application:flow}, the above information is present in the estimated group 1 across~age classes, for both the underlying and contributing causes. At age 0--1, this group includes a large number~of~causes~not typically associated with perinatal mortality, and then progressively fades at subsequent ages, further confirming the increasing heterogeneity of the causes--of--death~landscape~(which~is apparent also in terms of additional underlying and contributing causes joining the active groups).~Interestingly, Figure~\ref{fig:application:flow} highlights a  slight increase in the size of group 1 at age classes 80--90 and 90--100. While this pattern deserves further exploration, it may evidence selection effects in late life combined with susceptibility to fewer interactions among prevalent underlying and contributing~causes.

\vspace{-2pt}

Note that the change in size of group 1 is not only due to the emergence of causes exiting the inactive set at age 0--1. Rather, causes both appear and disappear with ages. For example, conditions associated with perinatal issues and malformations enter this group beyond the initial age class. Notably, while the most perinatal causes regarded as contributing factors are rapidly absorbed in group~1~by~the~second age class, many of these conditions (such as respiratory and cardiovascular disorders, and infections specific to the perinatal period) persist longer when regarded as underlying causes, joining group 1 only at later stages. This reflects the possibility that conditions emerging in the perinatal period may not turn out to be fatal at the same stage, but may persist into later life to become the cause that initiated the course of morbid events that led to death. In contrast, the same causes are unlikely to assume a contributing role in adulthood. Conversely, the causes exiting group 1 to join active clusters allow us to detect when a certain disease starts entering the mortality process and which already--active causes showcase its similar co--occurrence patterns. For example,~in~the~female~population, neoplasms of the respiratory organs, bones, and urinary tract exit group 1 in the transition between 0--1 years and 1--10 years, both for the underlying and contributing causes.
Other forms of neoplasms, such as those of the breast and female genital organs, enter the mortality process when transitioning from 1--10 to 10--20 years for the underlying causes, while this transition only manifests from 10--20 to 20--30 years in the contributing causes.
This suggests an asymmetry in the process, which may be due to the fact that early breast cancer is unlikely to be detected and thus is more likely to cause death as underlying factor rather than contributing one.
Other gender--specific phenomena~also~emerge,~particularly in older age groups for contributing causes. 
These differences, motivate further research on how contributing causes interact differently with underlying ones in males and females.

\begin{figure}[b]
    \centering
    \includegraphics[width=0.75\linewidth]{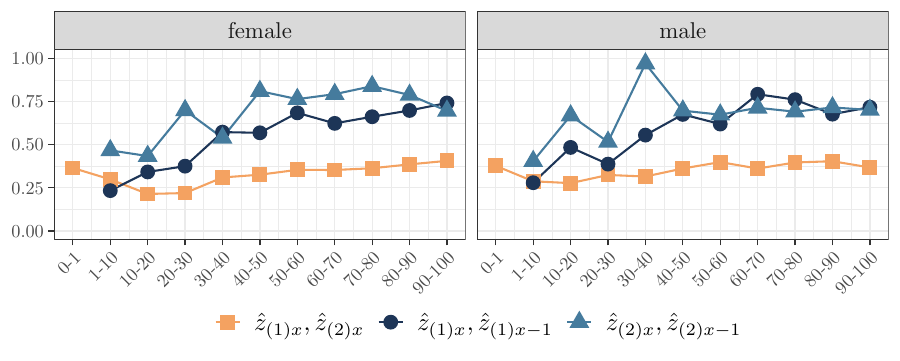}
        \vspace{-12pt}
    \caption{\footnotesize Normalized mutual information across ages between selected pairs (see the legend) of estimated~partitions.}
    \vspace{-13pt}
    \label{fig:application:nmi}
\end{figure}

To further expand the results in Figures~\ref{fig:intro:data_plot}--\ref{fig:application:flow}, we quantify in Figure~\ref{fig:application:nmi} the evolution of the similarity among the estimated partitions not only across consecutive age classes, but also with respect to contemporary group structures of the underlying and contributing causes at the same age. These~patterns~are~measured through $\textsc{nmi}(\hat{\bz}_{(1)x},\hat{\bz}_{(1)x-1})$, $\textsc{nmi}(\hat{\bz}_{(2)x},\hat{\bz}_{(2)x-1})$ and $\textsc{nmi}(\hat{\bz}_{(1)x},\hat{\bz}_{(2)x})$, respectively, for each  $x$, and provide additional empirical support in favor of our model. In particular, the evolution~of $\textsc{nmi}(\hat{\bz}_{(1)x},\hat{\bz}_{(2)x})$ evidences marked and stable differences among the partition structures of underlying and contributing causes. These asymmetries confirm the need for two separate partitions~and highlight how causes display  different group behaviors when acting as underlying or contributing. The importance of incorporating adaptive smoothness through age--specific persistence parameters $\alpha_{(1)x}$ and $\alpha_{(2)x}$ is instead confirmed by the evolution of $\textsc{nmi}(\hat{\bz}_{(1)x},\hat{\bz}_{(1)x-1})$ and $\textsc{nmi}(\hat{\bz}_{(2)x},\hat{\bz}_{(2)x-1})$ which showcase higher similarities among consecutive partitions of underlying and contributing causes, respectively, with an increasing trend that tends to  stabilize at elder ages.

\begin{figure}[t]
    \centering
    \includegraphics[width = 0.89\linewidth]{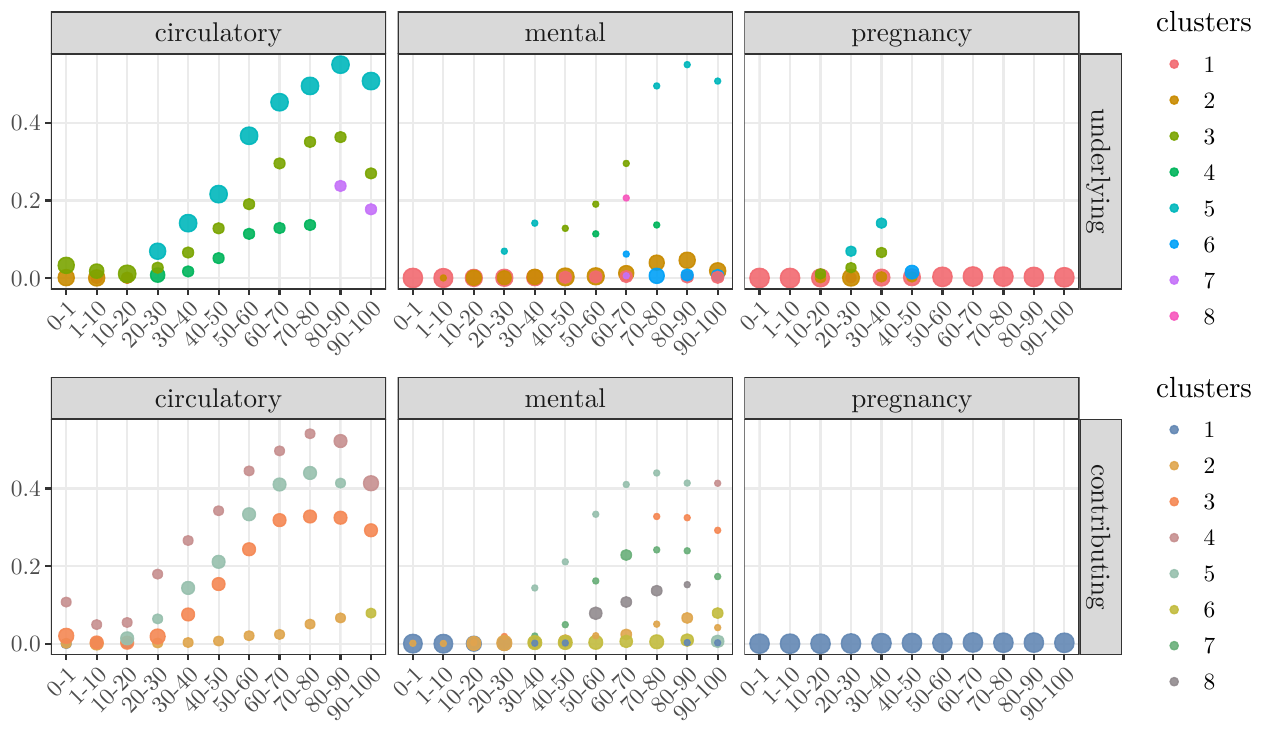}
    \caption{Group--specific marginal connectivity scores (\textsc{mcs}) over age classes for underlying (top) and contributing~(bottom) causes in the female population, stratified by three macro--categories (circulatory, mental, and pregnancy). Point size is proportional to the number of causes in each macro--category belonging to a given group.}
    \label{fig:application:intensity}
\end{figure}

While Figures~\ref{fig:intro:data_plot}, \ref{fig:application:flow} and \ref{fig:application:nmi} inform on the overall structure and transitions of the inferred groups,~Figure~\ref{fig:application:intensity} summarizes the relevance of such groups in terms of co--occurrence strengths and the associated composition with respect to three macro--categories of interest, namely, circulatory system conditions, mental health, and pregnancy--related causes. These macro--categories appear in the different panels of  Figure~\ref{fig:application:intensity}, where each point corresponds to a group and its size is proportional to the ratio between the number of causes from the macro--category analyzed that have been allocated to that group and the total number of causes in such a macro--category. Hence, points of large size correspond to groups absorbing most of the causes of death from the macro--category analyzed. The vertical axis measures instead the overall intensity displayed by the underlying (contributing) causes of a given group in co--occurring with those of all the contributing (underlying) clusters. These marginal connectivity scores (\textsc{mcs}) are defined as the averaged probability of observing a \texttt{present} or \texttt{frequent} interaction between the causes of a given underlying (contributing) group $h$ ($k$) and those belonging to a generic contributing (underlying) cluster $k=1, \ldots, K_x$ ($h=1, \ldots, H_x$). More specifically, at each age class~we~compute $\textsc{mcs}_{(1)hx} = (1/\hat{K}_x) \sum_{k = 1}^{\hat{K}_x} (\hat{\theta}_{hkx3} + \hat{\theta}_{hkx4})$, for $h=1, \ldots, \hat H_x$, and $\textsc{mcs}_{(2)kx} = (1/\hat{H}_x) \sum_{h = 1}^{\hat{H}_x} (\hat{\theta}_{hkx3} + \hat{\theta}_{hkx4})$, for $k=1, \ldots, \hat K_x$, where $\hat{\theta}_{hkx3}$ and $\hat{\theta}_{hkx4}$ are obtained as in \eqref{teta_plug_est}. The analysis of such measures~in~Figure~\ref{fig:application:intensity}, interestingly shows that pregnancy--related causes, while rare overall,  appear almost exclusively as underlying causes among women of childbearing age, with minimal presence as contributing factors. Mental health--related causes also display distinct patterns across panels, reflecting an evolving role in the progression of disease. As age grows, these conditions become increasingly common as contributing causes. In contrast, among women over 70, high prevalence as underlying causes is limited to organic mental disorders, including symptomatic conditions such as Alzheimer's and dementia, concentrated in cluster 5.
Circulatory system conditions are infrequent at younger ages but become more prominent in later stages for both roles. Still, underlying and contributing patterns diverge. For instance, five of the six circulatory causes (hypertensive, ischemic, pulmonary, cerebrovascular, and other heart diseases) co--cluster early (by ages 30–40) as underlying causes with high $\textsc{mcs}_{(1)hx}$. 
In contrast, contributing~roles are more fragmented, with these conditions spreading across~groups~of~varying~$\textsc{mcs}_{(2)hx}$.

\begin{figure}[t!]
    \centering
    \includegraphics[width=1.01\linewidth]{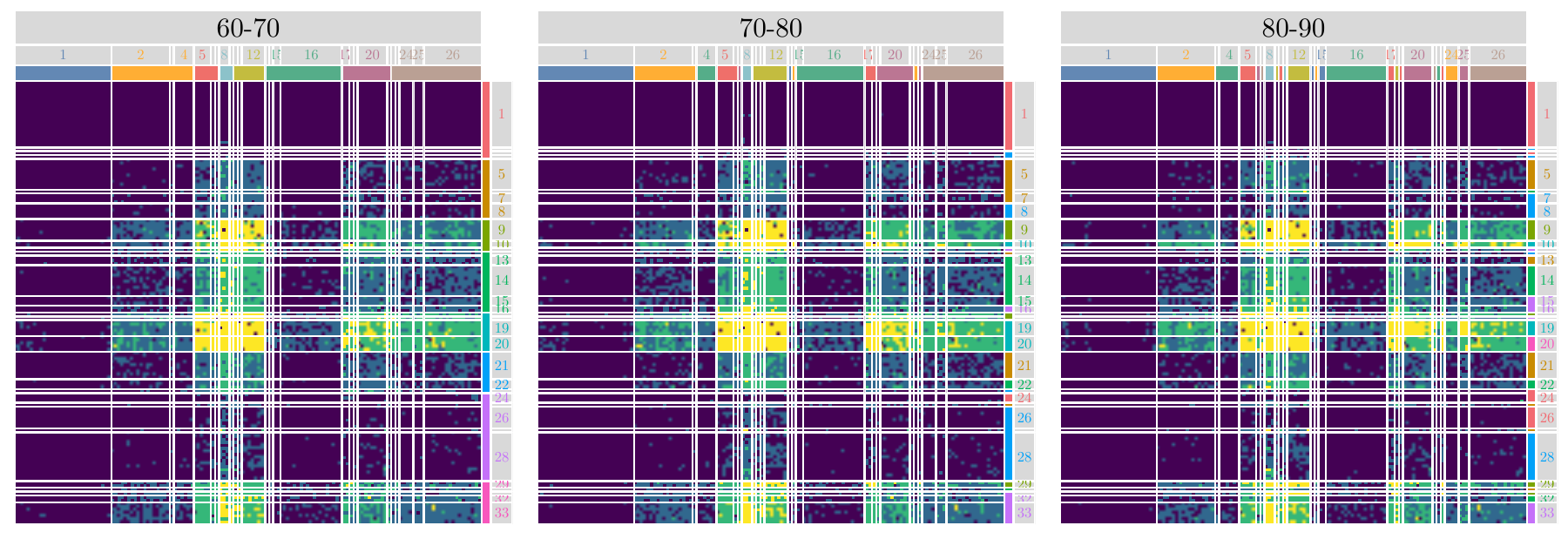}
    \vspace{-22pt}
    \caption{\footnotesize Adjacency matrices of the co--occurrence networks among underlying and contributing causes for~USA~female elder population. Rows and columns are re--ordered by estimated groups (colors) and {\em meet} clusters (lines--numbers). }
    \label{fig:application:meets}
\end{figure}

\begin{figure}[t!]
    \centering
    \includegraphics[width=1.01\linewidth]{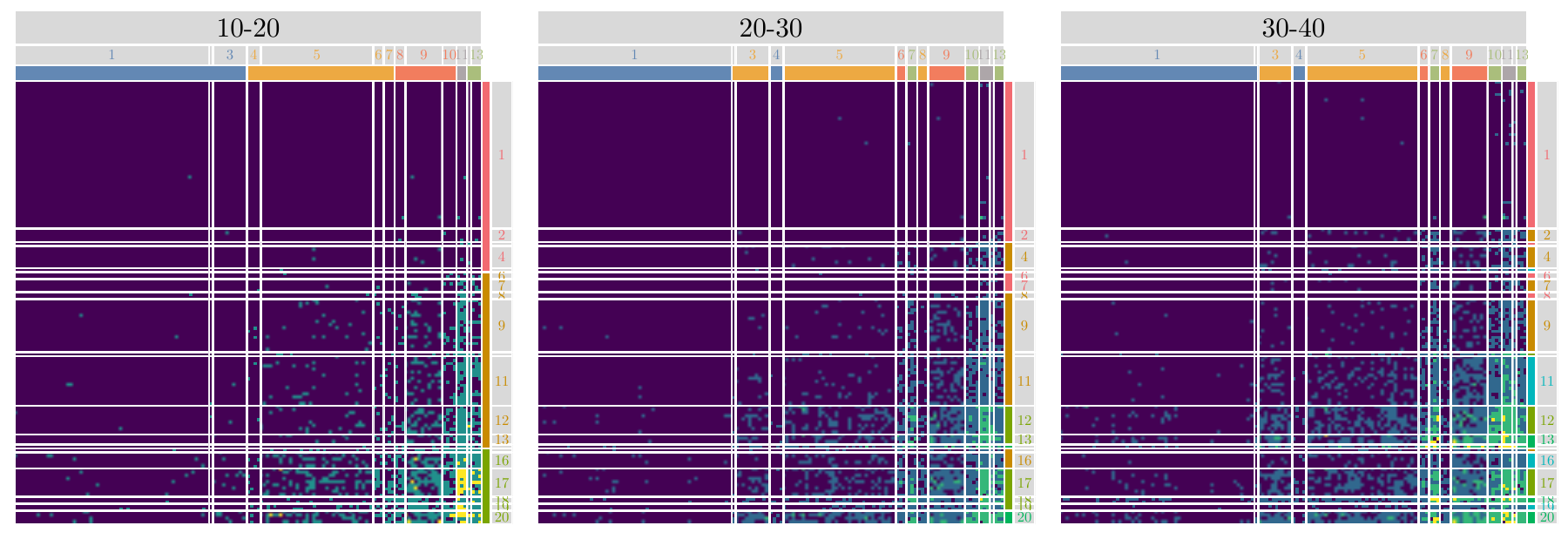}
        \vspace{-22pt}
    \caption{\footnotesize Adjacency matrices of the co--occurrence networks among underlying and contributing~causes~for~USA~male young population. Rows and columns are re--ordered by estimated groups (colors) and {\em meet} clusters (lines--numbers).}
    \label{fig:application:meets_male}
\end{figure}

\vspace{-2pt}
We conclude by exploring in more detail persistent groups and peculiar fragmentation patterns~that arise in the female and male population at elder (60--70, 70--80, 80--90) and adolescent/early--adulthood (10--20, 20--30, 30--40) ages, respectively. These analyses rely on the study of the meet clusters introduced in Section~\ref{subsec:posterior_summary} (see Figures~\ref{fig:application:meets}--\ref{fig:application:meets_male}) and highlight yet--unexplored patterns that  contribute~to~an improved understanding of the determinants underlying the recent growing mid--life mortality in the \textsc{usa} \citep[e.g.,][]{woolf2019life,mehta2020us,case2021deaths}. Focusing first on the female population, Figure~\ref{fig:application:meets} unveils a number of interesting meet clusters both~in~terms~of~composition and evolution. This is the case, for example, of the underlying meet clusters 9 (which includes respiratory and genitourinary conditions) and 10--11 (that comprise mental disorders and central nervous system neoplasms, respectively). While these three subsets of causes form a single group~in~the~60--70 class, at subsequent ages the meet cluster 9 eventually creates a separate~group~due~to~different~co--occurrence strengths with the contributing causes in  the meet cluster 20 (which~consists~of~severe~conditions such as lymphoid, breast and female genital neoplasms, and lung diseases caused by external agents). Similarly, the underlying meet clusters 19 and 20 (circulatory system causes and neoplasms, respectively) initially belong to the same group, due to similar interaction patterns with contributing causes. By age 80--90, however, the weakening of co--occurrences between the causes of the underlying meet cluster 20 and those of the contributing meet clusters 20 and 25 is in contrast with the strengthening of the interactions showcased by the underlying  meet cluster 19, thereby yielding to a fragmentation in two separate groups. Notice that, although modest in size, the  contributing meet cluster 25 (extrapyramidal and movement disorders, osteoarthritis and inflammatory polyarthropathies and other osteopathies), shows a very distinct pattern in the transition from the 70--80 to the 80--90~age class, as it separates from the adjacent meet clusters, primarily due to a strengthening of the co--occurrences~with causes from the underlying meet cluster 19. Interestingly, such a shift might signal~an~onset of age--related mobility dysfunctions contributing to death by aggravating circulatory conditions.

We now turn our attention to the analysis of the meet clusters for the male population  in those age classes that have witnessed peculiar mortality increments in \textsc{usa} over the recent years. As clarified~in Figure~\ref{fig:application:meets_male}, although the determinants of such increments have generated  a debate around different~views \citep[][]{woolf2019life,mehta2020us,case2021deaths}, a detailed analysis of the meet clusters and modules inferred by our model may~lead to a consensus among such views. 

Focusing first on the contributing causes, it is possible to recognize the meet cluster~11~as~a~highly--active one in the age class 10--20. This cluster mostly comprises various forms of heart and acute~respiratory diseases that often arise from infectious phenomena common to young ages. In fact, at early--adulthood stages, the meet cluster 11 becomes progressively~less active, whereas the contributing meet clusters 7 and 10 start displaying more intense interactions. Interestingly, these clusters include causes such as mental and behavioral disorders due to psychoactive substance use, diabetes mellitus, hypertensive diseases, diseases of liver, metabolic disorders, ischemic heart diseases, pulmonary heart disease and diseases of pulmonary circulation. This is an important finding which clarifies that, although current literature views cardiovascular diseases \citep[][]{mehta2020us} and ``death~of~despair''~type~causes \citep[][]{woolf2019life,case2021deaths} as alternative explanations of the recent increments in \textsc{usa} mid–life mortality, in fact, such causes cluster together, and hence, should not be treated as alternatives to each other, but rather as closely related and jointly contributing to overall mortality patterns. Such a result is further reinforced by the analysis of the underlying meet clusters in Figure \ref{fig:application:meets_male}, which display high co--occurrences for the meet clusters from 16 to 20 in  the age class 10--20. Similarly to the contributing meet cluster 11, these groups mainly encompass conditions related to respiratory diseases (e.g., flu, pneumonia, chronic lower respiratory diseases),~and~further~contain nervous system disorder (e.g., episodic and paroxysmal disorders, cerebral palsy and other paralytic syndromes, other disorders of the nervous system) along with neoplasms (e.g., musculoskeletal, eye, brain, central nervous system and endocrine gland cancers). Conversely, in populations aged 20--30 and~30--40,~a~second group of meet clusters (12, 13, 14) emerges through a transition from a low--interactions group~(orange) to the more active one (green). Interestingly, such a subset displays increasingly--strong co--occurrences with the contributing meet clusters 7 and 10, and is composed by a similar combination of cardiovascular and ``death of despair'' type causes. This result further strengthens the fact that the recent increments in \textsc{usa} mid–life mortality might arise from complex joint interactions among stochastically equivalent groups of underlying and contributing causes that appeared as alternatives to each other~in previous analyses focused only on the underlying cause \citep[see, e.g.,][]{woolf2019life,mehta2020us,case2021deaths}, rather than~on~multiple~interacting~ones.

\vspace{-10pt}

\section{Conclusions and Future Research}\label{sec_6}
\vspace{-10pt}
Motivated by the importance of achieving an in--depth understanding of the core structures regulating cause--of--death networks across ages, and by the lack of a suitable statistical model capable of inferring such structures, we designed in Sections~\ref{sec_2} and \ref{sec_3} an innovative stochastic block~model~for~sequences~of categorically--weighted  directed networks. Consistent with empirical evidence from real--data applications, such a model accounts for modular interactions among groups of causes, and crucially allows for two separate partition structures on the rows and columns of the asymmetric adjacency matrices, corresponding to underlying and contributing causes of death, respectively. Under a Bayesian approach to inference, these partitions are assigned flexible priors that induce smooth transitions in the group structures across age classes, further informed by external macro--classifications of death causes.  

The simulation studies in Section~\ref{sec_4} confirm that the proposed model is superior to state--of--the--art alternatives not only in realistic settings aligned with the motivating cause--of--death application, but also in simpler undirected networks scenarios favorable to currently--available formulations. This~motivates an extensive use of the proposed model in a wide variety of situations where multiple directed (or undirected) networks are observed across a temporal index. The application to \textsc{usa} causes--of--death networks in Section~\ref{sec_5} further confirms this point by showcasing the potential of the proposed model in unveiling fundamental group structures among underlying and contributing causes that were hidden from classical \textsc{mcod} and single--cause analyses. Besides the potential policy implications, the evolution and composition of such groups plays also an important role in the understanding of modern mortality trends, such as the recent increment of mid--life mortality in the \textsc{usa}.

\vspace{-2pt}

Interesting directions for future research emerge from our contribution. A natural one is to include further dimensions within our model, such as calendar years, geographical units (e.g., \textsc{usa} states), and others. This would require the design of a joint prior for random partitions that induces dependence not only between consecutive age classes, but also across contiguous calendar years and spatially--close units. Representations of this type could be found in the ongoing literature on Bayesian nonparametric priors relying on notions of separate exchangeability \citep{rebaudo2021separate}, conditional~partial~exchangeability \citep{franzolini2023conditional} and combinations of these two notions. Any advancement along these lines could be inherited within our formulation and would also help in the design of a joint model for the male and female populations that borrows information among the associated partitions.

\vspace{-2pt}
Besides the above methodological extensions, it shall be emphasized that the proposed model~naturally applies also to (rectangular) bipartite networks encoding interactions (possibly varying with~a~temporal index) among two different sets of entities (e.g., users and items interactions across years~in~recommender systems' applications), thereby motivating further exploration of the potential of our construction also in these settings. Additional empirical analyses are also of interest within the motivating causes--of--death application to assess whether the inferred group structures are robust to changes in the way through which interactions are defined. For example, as anticipated in Section~\ref{sec_1.D}, one option could be to replace the discretized co--occurrence strengths we employ, with normalized versions of the pairwise co--occurrence counts accounting for the overall degree of appearance of the involved causes in death certificates \citep[][]{hidalgo2009dynamic,chmiel2014spreading,fotouhi2018statistical}. Preliminary analyses show that the group structures we infer align also with the block patterns induced by this measure.
\vspace{-5pt}

\subsection*{Acknowledgments}
\vspace{-7pt}
This research is supported by the MUR–PRIN 2022 project CARONTE (Prot. 2022KBTEBN), funded by the European Union -- Next Generation EU, Mission 4, CUP: J53D23009400001. 

\vspace{4pt}
\let\oldbibliography\thebibliography
\renewcommand{\thebibliography}[1]{\oldbibliography{#1}
	\setlength{\itemsep}{5pt}} 

\spacingset{1.39}
\begingroup
\fontsize{11pt}{11pt}\selectfont
\bibliographystyle{JASA}
\bibliography{biblio.bib}
\endgroup

\end{document}